\documentclass[preprint2]{aastex61}
\usepackage[utf8]{inputenc}

\begin{document}
\sloppy


\title{Main-belt asteroids in the K2 Uranus field}

%% Use \author, \affil, plus the \and command to format author and affiliation 
%% information.  If done correctly the peer review system will be able to
%% automatically put the author and affiliation information from the manuscript
%% and save the corresponding author the trouble of entering it by hand.
%%
%% The \affil should be used to document primary affiliations and the
%% \altaffil should be used for secondary affiliations, titles, or email.

%% Authors with the same affiliation can be grouped in a single

\correspondingauthor{L. Moln\'ar}
\email{molnar.laszlo@csfk.mta.hu}

\author{L.~Moln\'ar}
\affiliation{Konkoly Observatory, MTA CSFK, Konkoly Thege Mikl\'os \'ut 15-17, H-1121 Budapest, Hungary}

\author{A.~P\'al}
\affiliation{Konkoly Observatory, MTA CSFK, Konkoly Thege Mikl\'os \'ut 15-17, H-1121 Budapest, Hungary}
\affiliation{E\"otv\"os Lor\'and University, P\'azm\'any P\'eter s\'et\'any 1/a, H-1117 Budapest, Hungary}

\author{K.~S\'arneczky}
\affiliation{Konkoly Observatory, MTA CSFK, Konkoly Thege Mikl\'os \'ut 15-17, H-1121 Budapest, Hungary}

\author{R.~Szab\'o}
\affiliation{Konkoly Observatory, MTA CSFK, Konkoly Thege Mikl\'os \'ut 15-17, H-1121 Budapest, Hungary}

\author{J.~Vink\'o}
\affiliation{Konkoly Observatory, MTA CSFK, Konkoly Thege Mikl\'os \'ut 15-17, H-1121 Budapest, Hungary}

\author{Gy.~M.~Szab\'o}
\affiliation{ELTE Gothard Astrophysical Observatory, 9704 Szombathely, Szent Imre herceg \'utja 112, Hungary}
\affiliation{Konkoly Observatory, MTA CSFK, Konkoly Thege Mikl\'os \'ut 15-17, H-1121 Budapest, Hungary}

\author{Cs.~Kiss}
\affiliation{Konkoly Observatory, MTA CSFK, Konkoly Thege Mikl\'os \'ut 15-17, H-1121 Budapest, Hungary}

\author{O.~Hanyecz}
\affiliation{Konkoly Observatory, MTA CSFK, Konkoly Thege Mikl\'os \'ut 15-17, H-1121 Budapest, Hungary}
\affiliation{E\"otv\"os Lor\'and University, P\'azm\'any P\'eter s\'et\'any 1/a, H-1117 Budapest, Hungary}

\author{G.~Marton}
\affiliation{Konkoly Observatory, MTA CSFK, Konkoly Thege Mikl\'os \'ut 15-17, H-1121 Budapest, Hungary}

\author{L.~L.~Kiss}
\affiliation{Konkoly Observatory, MTA CSFK, Konkoly Thege Mikl\'os \'ut 15-17, H-1121 Budapest, Hungary}
\affiliation{ELTE Gothard Astrophysical Observatory, 9704 Szombathely, Szent Imre herceg \'utja 112, Hungary}
\affiliation{Sydney Institute for Astronomy, School of Physics A28, University of Sydney, NSW 2006, Australia}

\begin{abstract}
We present the K2 light curves of a large sample of untargeted Main Belt asteroids (MBAs) detected with the \textit{Kepler} space telescope. The asteroids were observed within the Uranus superstamp, a relatively large, continuous field with low stellar background designed to cover the planet Uranus and its moons during Campaign 8 of the K2 mission. The superstamp offered the possibility to obtain precise, uninterrupted light curves of a large number of MBAs and thus to determine unambiguous rotation rates for them. We obtained photometry for 608 MBAs, and were able to determine or estimate rotation rates for 90 targets, of which 86 had no known values before. In an additional 16 targets we detected incomplete cycles and/or eclipse-like events. We found the median rotation rate to be significantly longer than that of the ground-based observations indicating that the latter are biased towards shorter rotation rates. Our study highlights the need and benefits of further continuous photometry of asteroids.
\end{abstract}

\keywords{techniques: photometric, minor planets, asteroids: general}



 \section{Introduction} \label{sec:intro}
The \textit{Kepler} space telescope has single-handedly revolutionized our understanding of exoplanets \citep[see, e.g.,][]{borucki2010,batalha2011-k11b,coughlin2016-dr24,kane2016-hzplanets}, matured the field of asteroseismology \citep{huber-redgiants,chaplin-ms-stars,sa-hosts}, and revitalized the interest in classical variable stars \citep{gilliland-2010,molnar-2016} during the last few years. The original mission was aimed towards the vicinity of the northern Ecliptic pole, allowing year-long, continuous observations \citep{borucki-2016}. However, after four years of operation, and the loss its second reaction wheel, \textit{Kepler} could not maintain the required precision in attitude control any longer. But the telescope was soon repurposed to observe along the Ecliptic plane in the K2 mission, in an attitude that minimized the uncompensated motion of the spacecraft \citep{howell-k2}.

The change from the vicinity of the Ecliptic pole to the plane resulted in a new, step--and--stare observation mode. The K2 mission is organized into campaigns that usually last 70--80 days. The new fields opened up the possibility to observe targets that the original mission lacked, such as young stars and nearby open clusters. Moreover, the K2 mission also brought various Solar System objects (SSOs) into view, widening the scope of the mission with planetary science applications.

Initial studies showed that the observations of \textit{Kepler} can be used to extract light curves of moving objects such as MBAs \citep{szabo-mba-initial,berthier-2016}. Since then, our group has developed a robust pipeline to identify and measure various classes of SSOs in the K2 observations and published a series of papers on trans-Neptunian objects \citep[TNOs,][]{pal-2015,or10}, the moon Nereid \citep{kiss-nereid}, as well as on Trojans and MBAs \citep{trojan,szabo-mba}. Independent studies also looked at the (intrinsic and reflected) light variations of the planet Neptune \citep{neptune-k2,gaulme,Rowe2017}.

\textit{Kepler} observations of asteroids may provide us with continuous light curves that surpass the length of ground-based measurements and thus can break the ambiguity in the determination of rotational periods caused by daily aliases. The results from the M35 and Neptune fields of the K2 mission hinted that the current sample of asteroid rotation periods may be biased towards shorter periods, although the K2 sample size was quite small \citep{szabo-mba}. Also, the median orbital period of known binary asteroids is in the order of one day that clearly hinders the detection of their photometric variation. Observational biases against long-period, low-amplitude asteroids also affect the statistics of various derived physical properties such as thermal inertia of the asteroid population \citep{marciniak}. Better statistics about the Solar System population also helps us to put various properties (such as color, rotation period and state, etc.) of interstellar asteroids like 1I/`Oumuamua into proper context \citep[see, e.g.,][]{meech2017,jewitt2017,bannister2017,fraser2017}. 

In this paper we set out to expand on this research, utilizing a superstamp that is larger than the Neptune field, but has a similarly low stellar background, unlike the dense M35 field. The planet Uranus was observable during K2 Campaign 8, and the planet and some of its moons were covered by pixel tiles that formed a continuous mosaic. An analysis of five distant, irregular moons has been published in a separate paper \citep{uranus-moons}. Here we present the analysis of the asteroids that were detected by the \textit{Kepler} space telescope. Section \ref{sec:data} describes the steps of reducing and analyzing the observations; in Sect.~\ref{sec:results} we show the various results and conclusions we derived; Sect.~\ref{sec:out} Appendix \ref{sec:app} contain a short outlook and a collection of the light curve plots.

\section{Data analysis}
\label{sec:data}

\begin{figure}
%\figurenum{5}
\epsscale{1.18}
\plotone{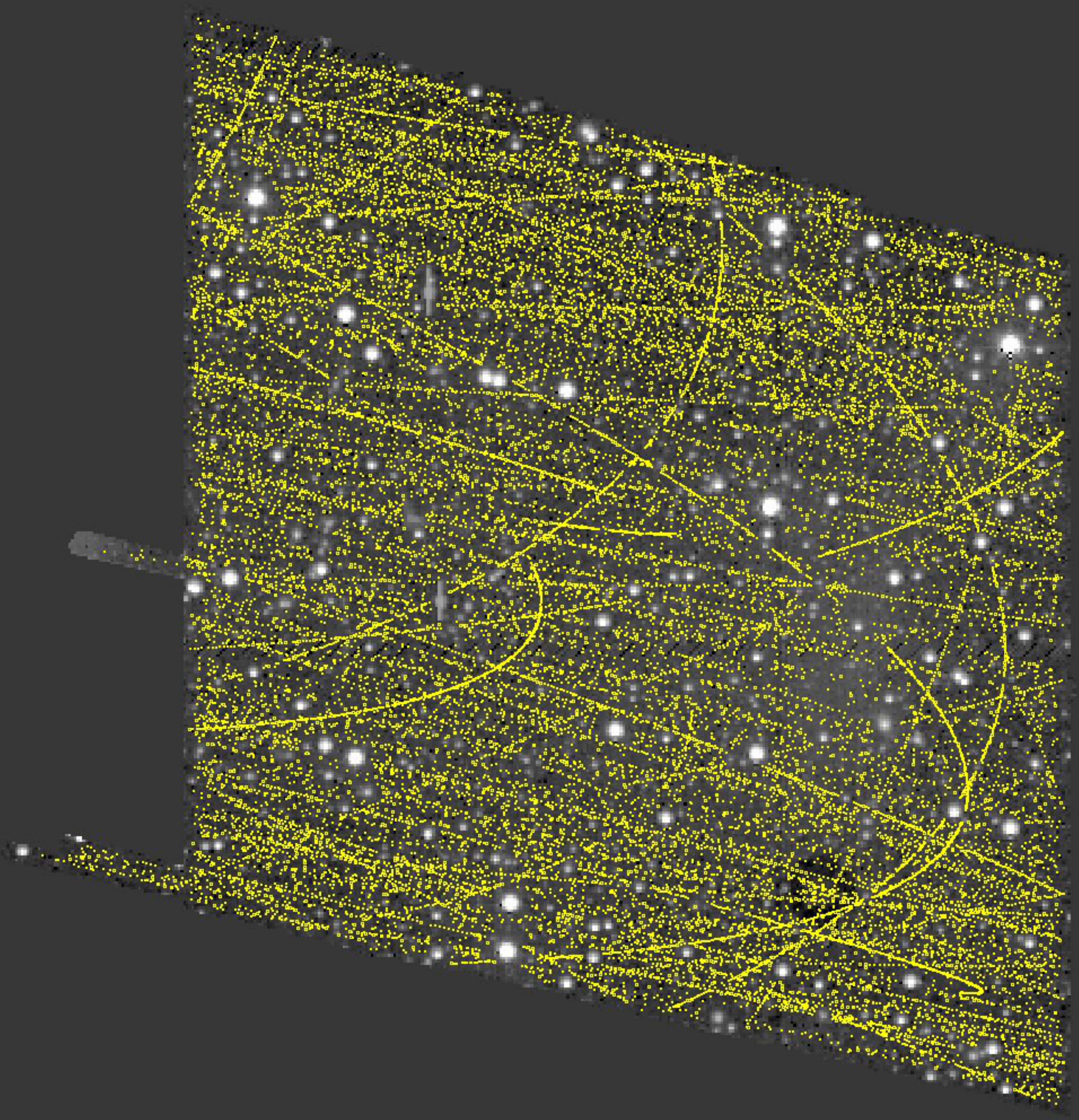}
\caption{The Uranus superstamp from the K2 mission and the tracks of all known asteroids across it. The white points are the background stars, while the yellow curves show the apparent tracks of the asteroids.}
\label{fig:tracks}
\end{figure}

The Uranus superstamp was observed via the K2 Guest Observer proposals GO8034 (PI: Jason F.\ Rowe), and GO8039 (PI: Andr\'as P\'al). Although parts of the superstamp were observed in short cadence mode too, we used long cadence data only (29.4 min integration time). None of the asteroids were proposed beforehand, therefore our sample consists of untargeted observations of a random population. This superstamp contains roughly $10^5$ pixels in a trapezoid shape covering an area of $21^\prime\times21^\prime$, plus the extensions corresponding to the tracks of the moons Setebos and Sycorax (see Fig.~\ref{fig:tracks}). The observations lasted from 2016 January 04.55 until March 23.28 (BJD 2457392.1--470.0), without interruptions. Overall, 3814 cadences were collected. 

As a comparison, our study is broadly similar to the Thousand Asteroid Light Curve Survey (TALCS) that used a larger telescope, the 3.5~m CFHT, a wider field--of--view, but obtained much sparser photometry a shorter time span of two weeks \citep{TALCS}.

\subsection{Source identification and photometry}
We processed the K2 data in the same way as in our previous works, utilizing the FITSH\footnote{\url{https://fitsh.net/}} software package \citep{fitsh} and our external scripts. Briefly, we created mosaic images covering the entire superstamp from the individual Target Pixel Files (TPFs) that contain the time series of small neighboring pixel areas selected for observation. We then derived the astrometric solutions for the mosaic images, using the USNO-B1.0 catalog, where the K2 full-frame images from the campaign were exploited as initial hints for the source cross matching. In the following step, we registered the images into the same reference system, and subtracted a median image from each image. This median image was created from a selection of individual images that did not include the halo and image of Uranus in them. 

Potential asteroid targets were identified by our own EPHEMD tool that is able perform searches for a given sky cone and time range simultaneously. We identified 674 asteroids that likely entered the mosaic image. Then we used EPHEMD to generate ephemerides for the selected asteroids and computed their positions for each individual cadence. We then applied aperture photometry to these positions. Since the proper motions of the asteroids are fast enough to smear their images over an LC integration, we used elongated apertures for the photometry, based on the directions and apparent velocities of the targets. The same technique was used to measure the brightness of the Trojan and other main belt asteroids \citep{szabo-mba,trojan}.

The sharp images of the stars that were shifted to compensate the attitude changes of the telescope, create characteristic residuals in the differential images that could contaminate the photometry of asteroids. Therefore we filtered out the epochs when the scatter of the background pixels in the photometric annulus was high. We also filtered out epochs when the asteroid was seen inside the halo cast by Uranus. We then discarded light curves with less than 12 data points, resulting in a final sample of 608 asteroids. The tracks of these asteroids are displayed in Fig.~\ref{fig:tracks}.

\begin{figure}
%\figurenum{5}
\epsscale{1.20}1
\plotone{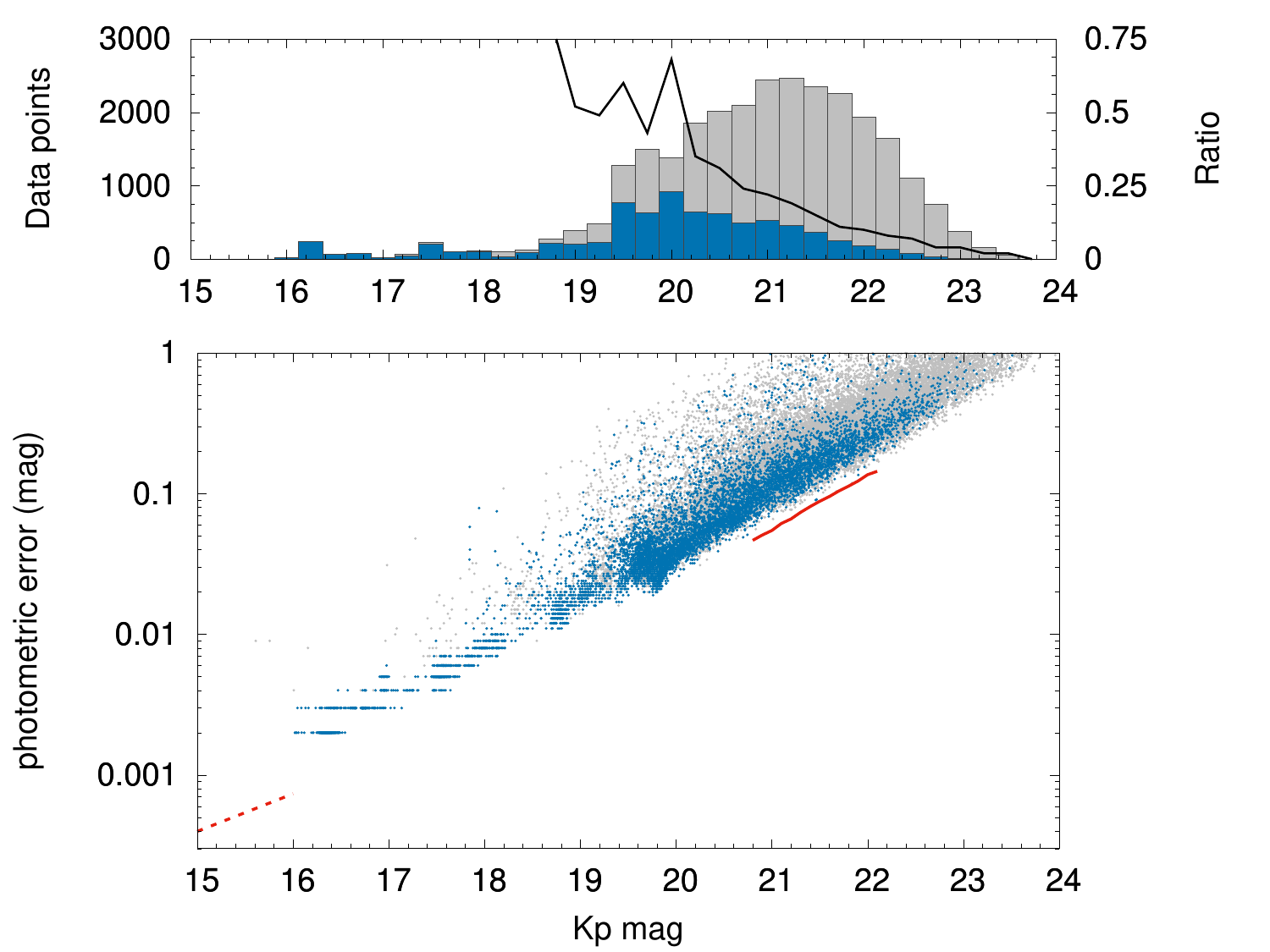}
\caption{Photometric errors of the individual data points showing the precision of the K2 measurements at the faint end. Grey points are all observations, blue points are those where we were able to determine rotation periods or clear temporal variations. The top plot shows the number of points at each 0.25 mag wide bin (grey: all, blue: with periods), and the ratio of the two measures is plotted with the black line. For comparison, the red dashed line follows the scaled Combined Differential Photometric Precision data (CDPP, \citealt{cdpp}), and the red solid line shows the errors for a faint RR Lyrae star from Leo IV \citep{molnar2015}. }
\label{fig:errors}
\end{figure}

We plotted the photometric uncertainty per data points in Fig.~\ref{fig:errors} according to the measured brightness values. The per-cadence photometric uncertainty values were derived on the shot noise of \textit{Kepler} and the estimated background noise. Beyond that, we decided to limit the photometric error to a conservative value of $10^{-3}$ or higher for bright targets, in order to account for any systematic variations that may be present in the data. The banded structure in Fig.~\ref{fig:errors} is caused by this limitation. The upper panel of Fig.~\ref{fig:errors} shows the ratio of all photometric points and those corresponding to the asteroids where we detected variations (see below). The figure indicates that \textit{Kepler} can detect the rotation signal is practically all asteroids above $Kp\sim 19$, but the ratio falls down to 25\% at 21~mag, and the faint limit is reached at approximately 22.5~mag. 

A sample of the data file containing measurements of the 608 asteroids is shown in Table~\ref{tab:lc}. Please note that slow trends were manually filtered from some targets during the analysis, but we have not removed those from the photometric data, in order to preserve any potential intrinsic signal. 

\begin{deluxetable}{lccc}
\tablecaption{Sample table of the photometry of main-belt asteroids observed in the Uranus superstamp during Campaign 8 of the K2 mission. The entire table is available online.\label{tab:lc}}
\tablehead{
\colhead{No./ID} &   \colhead{BJD (d)}  &  \colhead{USNO $R$ mag}  & \colhead{$\Delta R$ mag}}
\startdata
1570  &  2457433.6661  &  16.747  &  0.004\\
1570  &  2457433.6865  &  16.756  &  0.004\\
1570  &  2457433.7070  &  16.761  &  0.004\\
1570  &  2457433.9113  &  16.749  &  0.003\\
1570  &  2457433.9317  &  16.746  &  0.003\\
\multicolumn{4}{l}{\dots}\\
\enddata
\end{deluxetable}

\begin{figure*}
%\figurenum{5}
\epsscale{1.18}
\plotone{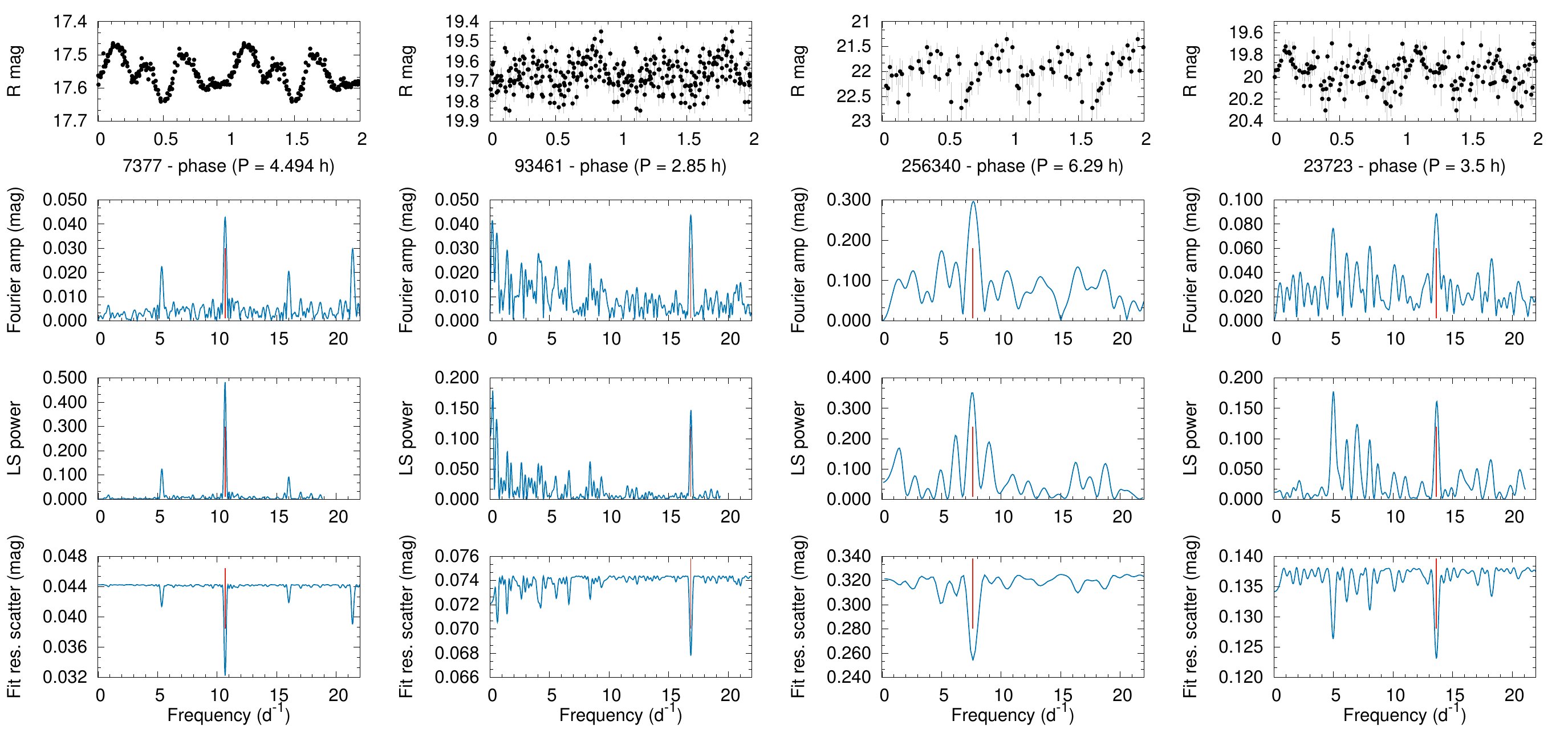}
\caption{Examples of the various period detection techniques: Fourier transform, Lomb-Scargle, and our own fit residual scatter algorithm. Asteroid (7737) shows a clear periodicity with additional peaks that describe the non-sinusoidal light curve. (93461) is a faint object that is affected by low-frequency noise but shows an obvious signal at higher frequencies. (256340) has a short data set, hence low frequency resolution. Finally, (23723) displays two possible periodicities: here we plotted the higher-frequency one that is more separated and is the most significant peak in 2 out of the 3 methods. Note that the plots display the phase curves twice to follow the variation when it folds at phase 1.0.}
\label{fig:spectra}
\end{figure*}

\vfill

\subsection{Period search}
We analyzed the obtained light curves with three different methods, namely, with Lomb-Scargle periodograms, fast Fourier transforms, and with our own residual minimisation-type algorithm. In the latter case, we fitted the data with the function $A+B\,{\rm cos}(2\pi f\Delta t)+C\,{\rm sin}(2\pi f\Delta t)$ where $f$ is the trial frequency, and $\Delta t = T-t$, where $T$ is the approximate center of the time series. We then search for the minimum in the dispersion of the residual curves for each value of $f$. We used the same method in previous studies too \citep{or10}. We inspected all light curves and the phase curves folded with the potential periods visually to select the candidates. A few examples are shown in Fig.~\ref{fig:spectra}. 

We tested the significances of the signals with multiple methods. We checked the signal-to-noise ratios (SNRs) of the FFT fits of the main frequency peaks, and set a lower limit of 3.5 in the vicinity of the peak. The noise was determined as a moving average of the spectrum using a 5 d$^{-1}$ window. We also tested the Gaussian nature of the noise distribution on the residual light curves after subtracting off the main frequency and its harmonics. The Anderson-Darling test of the light curve residuals showed that in the most cases, the residuals were compatible to the assumption of normality (76 and 71 asteroids at the 95\% and 98\% confidence levels, respectively).

\begin{figure}
%\figurenum{5}
\epsscale{1.1}
\plotone{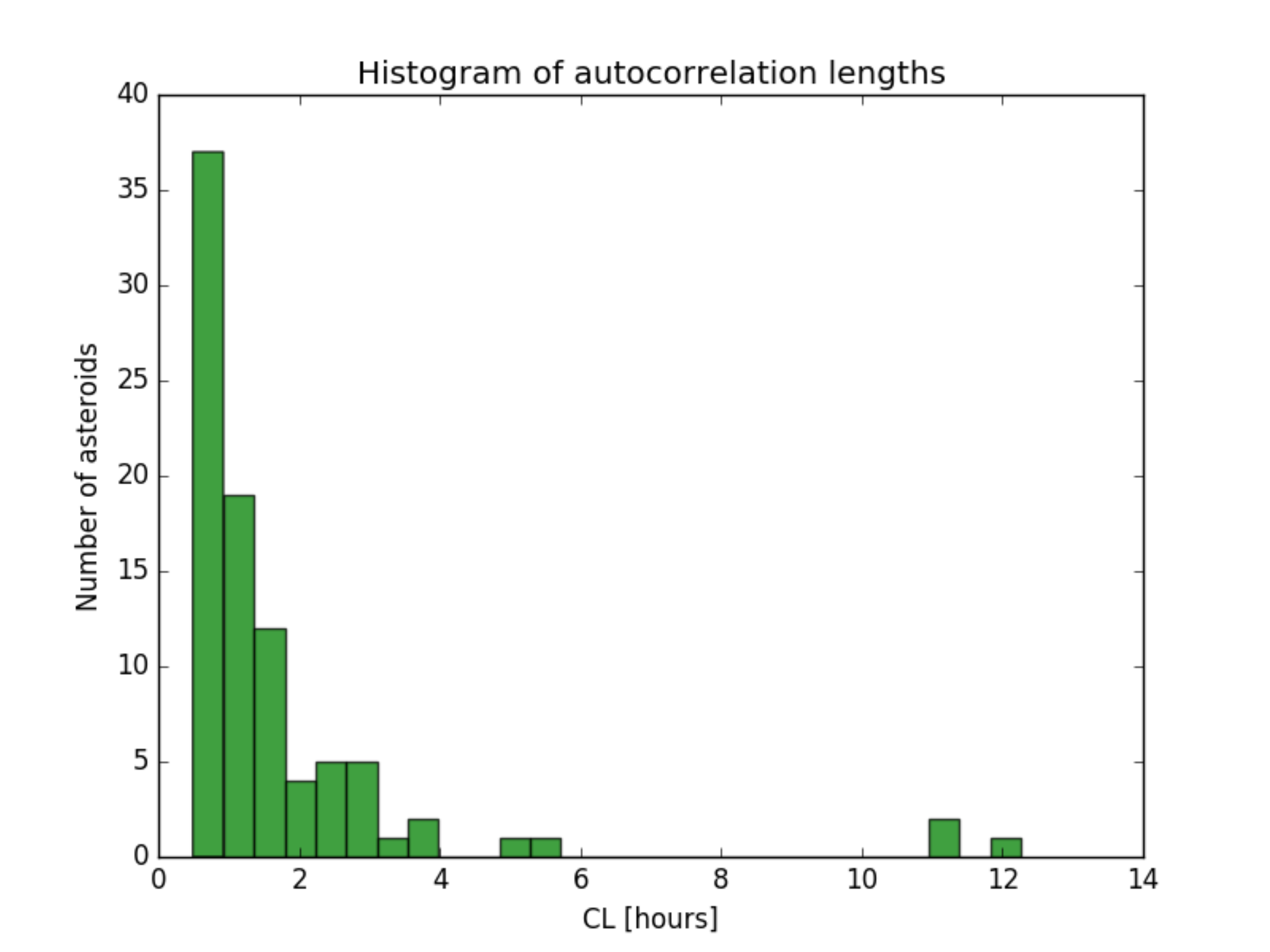}
\caption{Distribution of the autocorrelation lengths in the residual light curves. In almost all cases the values are shorter than 2 hours.}
\label{fig:autocorr}
\end{figure}

We further examined the possibility of significant autocorrelation in the residual light curves. We found that for the majority of our sample the correlation length in the residual light curve do not exceed 2 hours, as shown in Fig~\ref{fig:autocorr}, which is 4 times the cadence of the data sampling (0.5 hours). We also compared to the correlation length with the rotation periods and found that for the majority the CL/$P_{rot}$ ratio it is between 0.01 to 0.27, e.g. clearly shorter than the identified period. 

We examined the asteroids visually that failed one or more of the tests above (both the measurements and the residuals). We found that in some cases the residuals contain biases due to blends with stars along the path of motion that were not completely removed during photometry. (Given the shortness of the light curves, very few outlying points can already skew the distribution of the residual without significantly affecting the period determination.) In other cases the high-order harmonics were identified as a source of remanent signal, which were not removed when the residuals were calculated, and thus introduced autocorrelations and non-white components. Nevertheless, the majority of the flagged targets show high-amplitude asteroids with clean and unquestionable detections.
We identified only 7 cases where the detection was considered to be somewhat weak. We moved these asteroids into the second group of possible detections. 

Finally the visual inspection of the remaining light curves where periodicities were not found revealed 16 more asteroids with incomplete rotation cycles covered, or with eclipse-like features. 

\begin{figure}
%\figurenum{5}
\epsscale{1.18}
\plotone{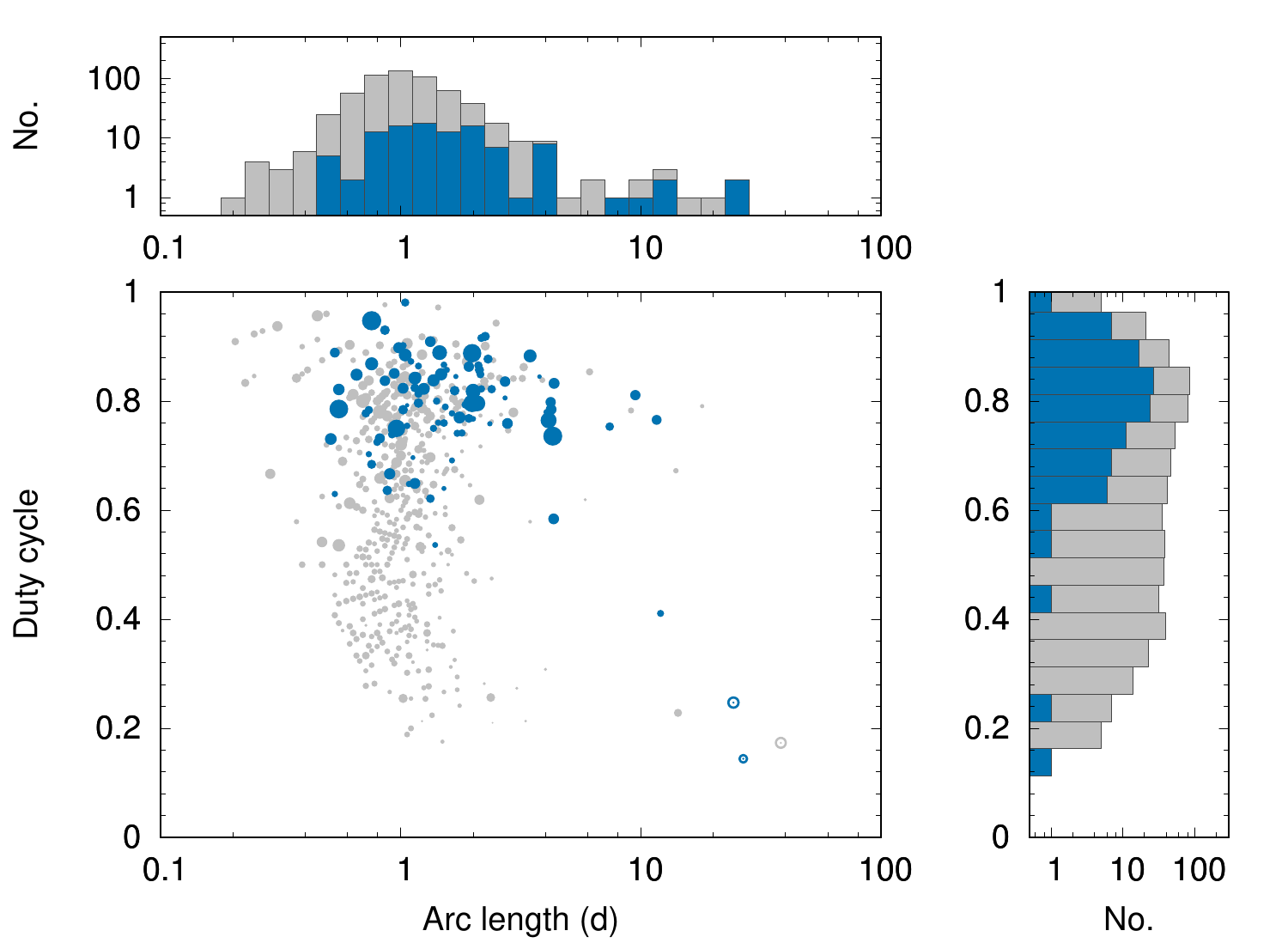}
\caption{Observed arc lengths and duty cycles. Grey points and bars are the entire sample, blue points are the asteroids where rotational variations were detectable.}
\label{fig:dutycycle}
\end{figure}

The duty cycles and arc lengths of the individual light curves are shown in Fig.~\ref{fig:dutycycle}. The median arc length is about 1.0 day which is about half of that of the M35 field \citep{szabo-mba}, but still roughly twice as long as any uninterrupted ground-based observation.

The duty cycle is defined as the fraction of useful data points in the light curve, 1.0 meaning no rejected K2 measurements. Here the distribution of duty cycles peaks around 0.8, thanks to the low number of stars: in contrast, the distribution for the M35 field peaked at 0.4. The sizes of the points in Fig.~\ref{fig:dutycycle} correspond to the average brightness of the asteroids, and their distribution shows that duty cycles are generally higher for brighter objects. Fainter asteroids are more susceptible to background variations, e.g., residuals of stellar PSFs and thus more data points were rejected from their photometry. Therefore the period search was hindered by both the lower photometric precision and the lower duty cycle for faint asteroids. 

We marked the asteroids with detectable variations in Fig.~\ref{fig:dutycycle} with blue points and bars. Most of them had duty cycles higher than 0.6 that agrees with our conclusions for the M35 field earlier \citep{szabo-mba}. Notably, we identified three asteroids -- (22438) 1996 HQ$_{19}$, (194109) 2001 SX$_{231}$, (359794) 2011 UF$_{198}$ -- that crossed the superstamp twice, both before and after they reached their stationary points. The gap between the two crossings lowered the duty cycles of these asteroids: we marked them with circles in Fig.~\ref{fig:dutycycle}. 

\section{Results}
\label{sec:results}
\subsection{New rotational periods}
We were able to conclusively determine the rotational periods of 79 asteroids, and derive potential periods for 11 more from the 608 targets we measured. Most of the light curves were symmetric and one-peaked: however, as we expect the light variations to arise primarily from the variations of the illuminated surface area of the body as it rotates, we adopted the double-peaked solutions in all cases. The rotation rates and photometric peak-to-peak amplitudes of these 90 asteroids are summarized in Tables~\ref{tab:rot1} and \ref{tab:uncrot}, while the lower limits for an additional 16 targets are listed in Table~\ref{tab:rot2}. We note that one of the asteroids, (89973) Aranyj\'anos was discovered by one of the authors as part of their ongoing Solar System astrometric program \citep{aranyjanos}. 

We compared our sample with the rotation parameters found in the Asteroid Lightcurve Database \citep{warner-ald}. Only six matches were found: in the cases of (2193) Jackson and (80969) 2000 DL$_{112}$ our rotational periods agree with earlier results. The K2 data for the two slow rotators (10936) 1998 FN$_{11}$ and (356766) 2011 UT$_{269}$ were too short to cover an entire cycle, but the light curves are consistent with the literature values. We did not find conclusive periods for the fast rotators (19289) 1996 HY$_{12}$ and (92018) 1999 VW$_{162}$ based on the K2 light curves, and folding with the published periods did not give conclusive results either. Numerical values are listed in Table~\ref{tab:rot3}. 

We also compared our list with the Database of Asteroid Models from Inversion Techniques \citep[DAMIT,][]{damit} to see if any of our light curves can be compared with synthetic light curves from shape models, but found no matches with that database. 

\begin{figure}
%\figurenum{5}
\epsscale{1.2}
\plotone{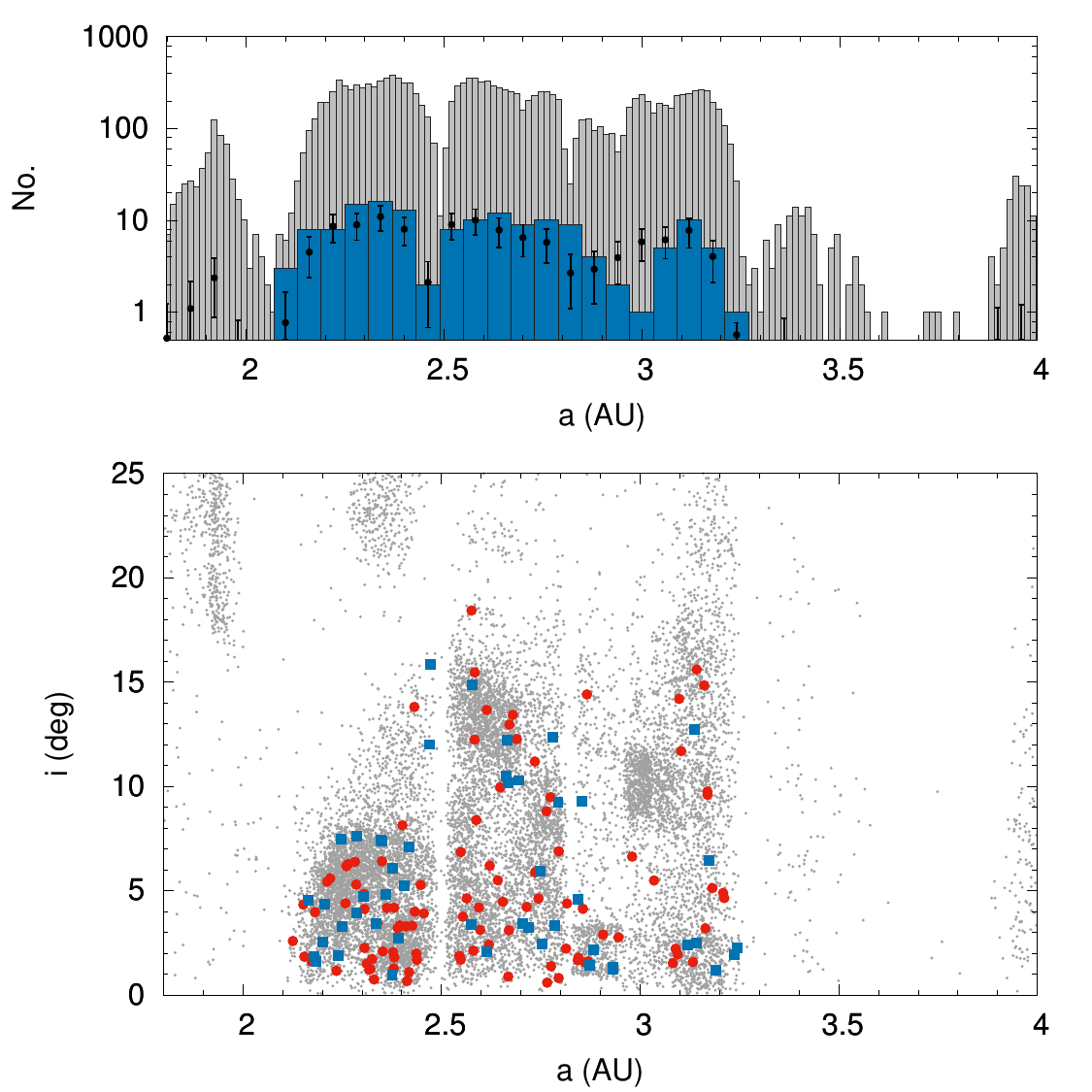}
\caption{Top: the histogram of the number of asteroids (grey: all, blue: K2 sample, black: simulated values without inclination preference). Bottom: orbital parameters of the K2 asteroids with detected rotational variations, compared to all asteroids with rotational parameters, by \citet{warner-ald}. Red points are asteroids from this study, blue points are from \citet{szabo-mba}. }
\label{fig:orbits}
\end{figure}

\begin{deluxetable*}{ccc|ccc|ccc}
\tabletypesize{\footnotesize}
\tablecaption{Rotation parameters of the asteroids from the Uranus superstamp. No./ID refers to the number or  provisional designation of the object.\label{tab:rot1}} 
\tablehead{
\colhead{No./ID} & \colhead{P (h)} & \colhead{A (mag)} & \colhead{ID} & \colhead{P (h)} & \colhead{A (mag)}& \colhead{No./ID} & \colhead{P (h)} & \colhead{A (mag)}}
\startdata
1878	&	$	2.68	\pm	0.07	$	&	$	0.10	\pm	0.01	$	&	84360	&	$	6.0 	\pm	0.4 	$	&	$	0.71	\pm	0.07	$	&	202122	&	$	5.40	\pm	0.07	$	&	$	0.78	\pm	0.03	$	\\
2193	&	$	4.8 	\pm	0.4 	$	&	$	0.24	\pm	0.03	$	&	88521	&	$	5.46	\pm	0.13	$	&	$	0.99	\pm	0.04	$	&	207030	&	$	2.67	\pm	0.08	$	&	$	0.03	\pm	0.01	$	\\
2255	&	$	9.78	\pm	0.16	$	&	$	0.41	\pm	0.02	$	&	91172	&	$	21.7	\pm	3.5 	$	&	$	0.13	\pm	0.04	$	&	209188	&	$	6.9 	\pm	0.7 	$	&	$	0.57	\pm	0.09	$	\\
5706	&	$	8.25	\pm	0.15	$	&	$	0.24	\pm	0.01	$	&	92871	&	$	12.4	\pm	0.5 	$	&	$	0.58	\pm	0.09	$	&	214304	&	$	11.7	\pm	0.7 	$	&	$	0.51	\pm	0.06	$	\\
6452	&	$	7.6 	\pm	0.5 	$	&	$	0.49	\pm	0.05	$	&	93177	&	$	7.34	\pm	0.24	$	&	$	1.05	\pm	0.05	$	&	229020	&	$	37.9	\pm	0.5 	$	&	$	0.95	\pm	0.04	$	\\
7737	&	$	4.49	\pm	0.20	$	&	$	0.08	\pm	0.01	$	&	93461	&	$	2.85	\pm	0.19	$	&	$	0.09	\pm	0.02	$	&	233123	&	$	33.0	\pm	3.3 	$	&	$	0.90	\pm	0.14	$	\\
10936	&	$	28  	\pm	4   	$	&	$	0.40	\pm	0.01	$	&	94314	&	$	31.0  	\pm	0.3 	$	&	$	0.49	\pm	0.02	$	&	234522	&	$	5.7 	\pm	0.9 	$	&	$	0.94	\pm	0.15	$	\\
13407	&	$	4.26	\pm	0.05	$	&	$	0.38	\pm	0.01	$	&	94988	&	$	5.43	\pm	0.05	$	&	$	0.55	\pm	0.02	$	&	238321	&	$	25.6	\pm	1.6 	$	&	$	0.21	\pm	0.03	$	\\
14379	&	$	2.7 	\pm	0.5 	$	&	$	0.05	\pm	0.01	$	&	98290	&	$	3.3 	\pm	0.4 	$	&	$	0.34	\pm	0.08	$	&	245568	&	$	7.60	\pm	0.26	$	&	$	0.76	\pm	0.05	$	\\
14527	&	$	6.01	\pm	0.25	$	&	$	0.88	\pm	0.07	$	&	101256	&	$ 5.1	\pm	0.6 	$	&	$	0.18	\pm	0.06	$	&	256340	&	$	6.3 	\pm	0.6 	$	&	$	0.56	\pm	0.11	$	\\
17715	&	$	6.16	\pm	0.08	$	&	$	1.04	\pm	0.03	$	&	103110	&	$	9.8 	\pm	2.5 	$	&	$	0.51	\pm	0.15	$	&	279947	&	$	6.6 	\pm	0.4 	$	&	$	0.44	\pm	0.07	$	\\
19425	&	$	9.1 	\pm	0.6	    $	&	$	0.19	\pm	0.02	$	&	108196	&	$	2.74	\pm	0.26	$	&	$	0.20	\pm	0.04	$	&	296879	&	$	4.96	\pm	0.18	$	&	$	0.61	\pm	0.06	$	\\
22478	&	$	21.6	\pm	1.9 	$	&	$	0.25	\pm	0.05	$	&	131712	&	$	25  	\pm	5   	$	&	$	0.70	\pm	0.15	$	&	301268	&	$	23  	\pm	3   	$	&	$	1.20	\pm	0.23	$	\\
25163	&	$	34.3	\pm	0.4 	$	&	$	0.43	\pm	0.01	$	&	140562	&	$	5.9 	\pm	0.4 	$	&	$	0.74	\pm	0.10	$	&	302685	&	$	22.0	\pm	1.2 	$	&	$	0.90	\pm	0.08	$	\\
27363	&	$	12.8	\pm	0.4 	$	&	$	0.09	\pm	0.01	$	&	145064	&	$	4.89	\pm	0.20 	$	&	$	0.45	\pm	0.03	$	&	345804	&	$	22.3	\pm	1.4 	$	&	$	0.57	\pm	0.11	$	\\
29923	&	$	17.9	\pm	2.0 	$	&	$	0.09	\pm	0.01	$	&	152876	&	$	4.10 	\pm	0.06	$	&	$	0.51	\pm	0.02	$	&	359794	&	$	53.5	\pm	1.6 	$	&	$	0.32	\pm	0.03	$	\\
33543	&	$	82  	\pm	10  	$	&	$	0.60	\pm	0.02	$	&	152926	&	$	43  	\pm	4   	$	&	$	0.70	\pm	0.14	$	&	387038	&	$	5.1 	\pm	0.8 	$	&	$	0.79	\pm	0.16	$	\\
34402	&	$	71.8	\pm	2.2 	$	&	$	0.43	\pm	0.03	$	&	156094	&	$	5.5 	\pm	0.8 	$	&	$	0.32	\pm	0.09	$	&	392778	&	$	8.7 	\pm	1.4 	$	&	$	0.88	\pm	0.22	$	\\
37460	&	$	6.4 	\pm	0.7 	$	&	$	0.60	\pm	0.06	$	&	156230	&	$	5.6 	\pm	0.6 	$	&	$	0.80	\pm	0.16	$	&	438341	&	$	6.7 	\pm	0.5 	$	&	$	0.69	\pm	0.11	$	\\
38963	&	$	10.60	\pm	0.11	$	&	$	0.60	\pm	0.02	$	&	165543	&	$	4.7 	\pm	0.3 	$	&	$	0.36	\pm	0.06	$	&	444569	&	$	12.0  	\pm	0.6 	$	&	$	1.2  	\pm	0.4 	$	\\
45086	&	$	5.0   	\pm	0.4 	$	&	$	0.41	\pm	0.05	$	&	169703	&	$	5.4 	\pm	0.9 	$	&	$	0.69	\pm	0.13	$	&	448497	&	$	12.1	\pm	1.0 	$	&	$	0.50	\pm	0.09	$	\\
50934	&	$	17.9	\pm	2.7 	$	&	$	0.53	\pm	0.12	$	&	171605	&	$	7.5 	\pm	2.1 	$	&	$	0.53	\pm	0.17	$	&	448647	&	$	12.2	\pm	0.8 	$	&	$	0.40	\pm	0.07	$	\\
51828	&	$	3.6 	\pm	0.4 	$	&	$	0.88	\pm	0.15	$	&	182196	&	$	6.7 	\pm	0.9 	$	&	$	0.10	\pm	0.03	$	&	449579	&	$	4.5 	\pm	0.4 	$	&	$	0.67	\pm	0.15	$	\\
58163	&	$	6.1 	\pm	0.8 	$	&	$	0.71	\pm	0.12	$	&	187014	&	$	9.3 	\pm	0.7 	$	&	$	0.41	\pm	0.10	$	&	484673	&	$ 12.9 \pm 1.5	$	&	$	0.60 	\pm	0.20	$	\\
59168	&	$	17.6	\pm	3.0 	$	&	$	0.60	\pm	0.12	$	&	194109	&	$	3.38 	\pm	0.07	$	&	$	0.13	\pm	0.01	$	&	2002 WY$_{28}$	&	$	6.88	\pm	1.80	    $	&	$	0.12	\pm	0.13	$	\\
80969	&	$	13.9	\pm	1.1 	$	&	$	0.82	\pm	0.14	$	&	199142	&	$	8.9 	\pm	1.0 	$	&	$	0.45	\pm	0.11	$	&							~	&					~	\\
83109	&	$	21.1	\pm	0.8 	$	&	$	0.49	\pm	0.04	$	&	200761	&	$	13.0  	\pm	1.4 	$	&	$	0.60	\pm	0.16	$	&							~	&					~	\\
\enddata
\end{deluxetable*}

\begin{deluxetable}{ccc}
\tablecaption{Asteroids with possible period detections. In three cases we list two different vaules, respectively.\label{tab:uncrot}}
\tablehead{\colhead{No.} & \colhead{P (h)} & \colhead{A (mag)} }
\startdata
23723	&	$	3.5	\pm	0.3	$	&	$	0.18	\pm	0.04	$	\\
--	&	$	9.6	\pm	0.4	$	&	$	0.15	\pm	0.05	$	\\
24066	&	$	11.1 	\pm	2.4	$	&	$	0.43	\pm	0.14	$	\\
31101	&	$	10.0	\pm	1.5	$	&	$	0.93	\pm	0.22	$	\\
60008	&	$	8.1	    \pm	1.6	$	&	$	0.16	\pm	0.04	$	\\
--		&	$	2.7	\pm	0.8	$	&	$	0.13	\pm	0.04	$	\\
89973	&	$	3.5	\pm	0.6	$	&	$	0.18	\pm	0.08	$	\\
127963	&	$	2.3	\pm	0.4	$	&	$	0.12	\pm	0.04	$	\\
140194	&	$	3.30	\pm	0.15	$	&	$	0.17	\pm	0.03	$	\\
--	&	$	3.00	\pm	0.18	$	&	$	0.13	\pm	0.03	$	\\
308977	&	$	6.7	\pm	0.6	$	&	$	0.33	\pm	0.07	$	\\
355994	&	$	7.2	\pm	0.7	$	&	$	0.24	\pm	0.07	$	\\
356102	&	$	20.4	\pm	0.5	$	&	$	0.62	\pm	0.22	$	\\
2005 QX$_{175}$	&	$	3.6	 \pm	0.9	$	&	$	0.54	\pm	0.19	$	\\
\enddata
\end{deluxetable}

\begin{deluxetable}{ccc|ccc}
\tabletypesize{\footnotesize}
\tablecaption{Asteroids with slow variations. (88347) 2000~HL$_{31}$ shows and eclipse, but no other variations \label{tab:rot2}}
\tablehead{\colhead{No.} & \colhead{P (h)} & \colhead{A (mag)} & \colhead{ID} & \colhead{P (h)} & \colhead{A (mag)}}
\startdata
1570	&	$>48.$	&	$>0.6$	&	92237	&	$>48.$	&	$>0.5$  \\
9611	&	$>48.$	&	$>0.25$	&	190345	&	$>24.$	&	$>1$	\\
16484	&	$>24.$	&	$>0.5$	&	282500	&	$>24.$	&	$>1$	\\
42992	&   $>36.$  &   $>0.7$  &	289668	&	$>48.$	&	$>1.2$	\\   
63486	&	$>24.$	&	$>0.1$	&	304953	&	$>48.$	&	$>0.5$	\\
69524	&	$>48.$	&	$>1.3$	&	309020	&	$>48.$	&	$>1.2$	\\
75722	&	$>24.$	&	$>0.5$	&	356766	&	$>24.$	&	$>1$	\\
88347   &	 ecl.    &	 --	    &	392619	&	$>48.$	&	$>1$	\\
\enddata
\end{deluxetable}

\begin{deluxetable}{cccc}
\tabletypesize{\footnotesize}
\tablecaption{Rotation parameters of the asteroids where literature values are available. \label{tab:rot3} }
\tablehead{
\colhead{No.} & \colhead{P (h)} & \colhead{A (mag)} & \colhead{Ref.}}
\startdata
2193	&	$4.79\pm 0.34$	&	0.24	&	this paper	\\
~		&	4.7541	&	0.24	&	\citet{Oey-2009}	\\
~		&	4.755	&	0.23	&	\citet{Wacak-2015} \\
80969	&	$13.9\pm 1.1$ 	&	0.82	&	this paper	\\
~		&	14.33	&	0.74	&	\citet{Chang-2015}\\
~		&	14.559	&	0.64	&	\citet{Wacak-2015}\\	
10936	&	$28\pm 4$	&	0.4	&	this paper		\\
~		&	17.3	&	0.35	&	\citet{Warner-2008}	\\
~		&	25.7	&	0.28	&	\citet{Warner-2011}	\\
356766	&	$>24.$	&	$>1$	&	this paper	\\
~		&	37.848	&	0.56	&	\citet{Wacak-2015}	\\
\enddata
\end{deluxetable}

Then we examined where the asteroids with detected rotation signals reside within the asteroid belt. It turns out that all 90+16 orbit within the Main Asteroid Belt, between $a=2.12$~AU and 3.24~AU ((289668) 2005~GN$_{137}$ and (17771) Elsheimer being the closest and farthest, respectively). Distribution of the K2 sample, including those by \citet{szabo-mba}, in the $a$--$i$ plane are shown in Fig.~\ref{fig:orbits}. The histogram at the top shows that the numbers of the K2 sample follow the distribution of other asteroids with orbital parameters (as well as the distribution of all known asteroids). The deviations from the expected numbers (with black symbols), and the lack of nearer asteroids can be attributed to the preference of low-inclination targets of the mission. 

\subsection{Peculiar light curves and eclipses}
Most K2 light curves show either rather simple variations that can be attributed to roughly ellipsoidal shapes, or are too uncertain to provide information about the shape of the asteroids. Some other light curves suggest contact binary bodies, with inverted U-shaped maxima, and deep, short, V-shaped minima in their light curves, in accordance with the model predictions \citep{Lacerda2007}. Contact binaries are elongated towards each other due to mutual gravitational forces and therefore their light curve amplitudes can be quite high \citep{leone1984}. The best candidates are asteroids (17715) 1997 WZ$_{39}$, (25163) Williammcdonald, (88521) 2001 QN$_{168}$, (93177) 2000 SP$_{103}$, (94988) 2001 YU$_{119}$, (229020) 2003 YU$_{68}$, (245568) 2005 UR$_{150}$, and (302685) 2002 TA$_{57}$. The K2 observations have phase angles less than 30$\degr$, so V-shaped minima with no or very short flat-bottomed sections are expected \citep{Lacerda2007}. These objects are therefore good candidates for further follow-up observations to confirm or disprove binarity. 

Beyond these, we also identified a few interesting cases during the visual inspection of the light curves and phase plots. Asteroids (7737)~Sirrah and (27363)~Alvancark show relatively complex light curves with multiple peaks. (7737)~Sirrah features four maxima and minima at different heights. The light variations of (27363)~Alvancark exhibit a long, plateau-like feature that is not completely flat and likely include additional, local minima and maxima, close to the detection limit of our data set. 

\begin{figure}
%\figurenum{5}
\epsscale{1.18}
\plotone{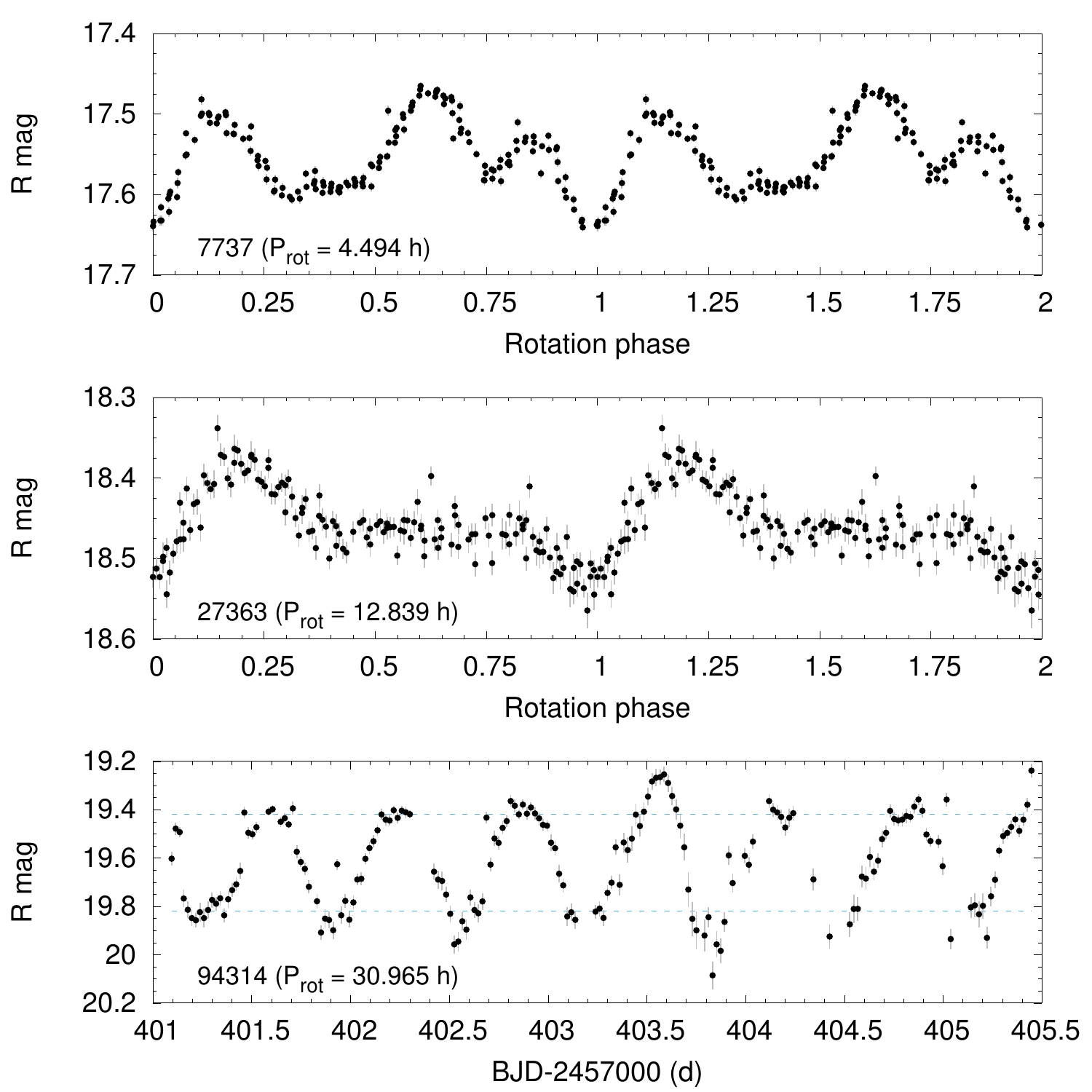}
\caption{Top: phase curves of two asteroids with multi-peaked light curves. (7737) Sirrah feature likely four different peaks per rotation. The light curve of (27363) Alvancark shows one dominant bright phase and a plateau-like feature that contains two smaller bumps. Bottom: the light curve (94314) 2000 AO$_{165}$, a possible tumbling or precessing asteroid that shows slightly changing cycles and one at BJD 2457403.5 with peculiarly large amplitude.  }
\label{fig:threepeaks}
\end{figure}

\begin{figure}
%\figurenum{5}
\epsscale{1.2}
\plotone{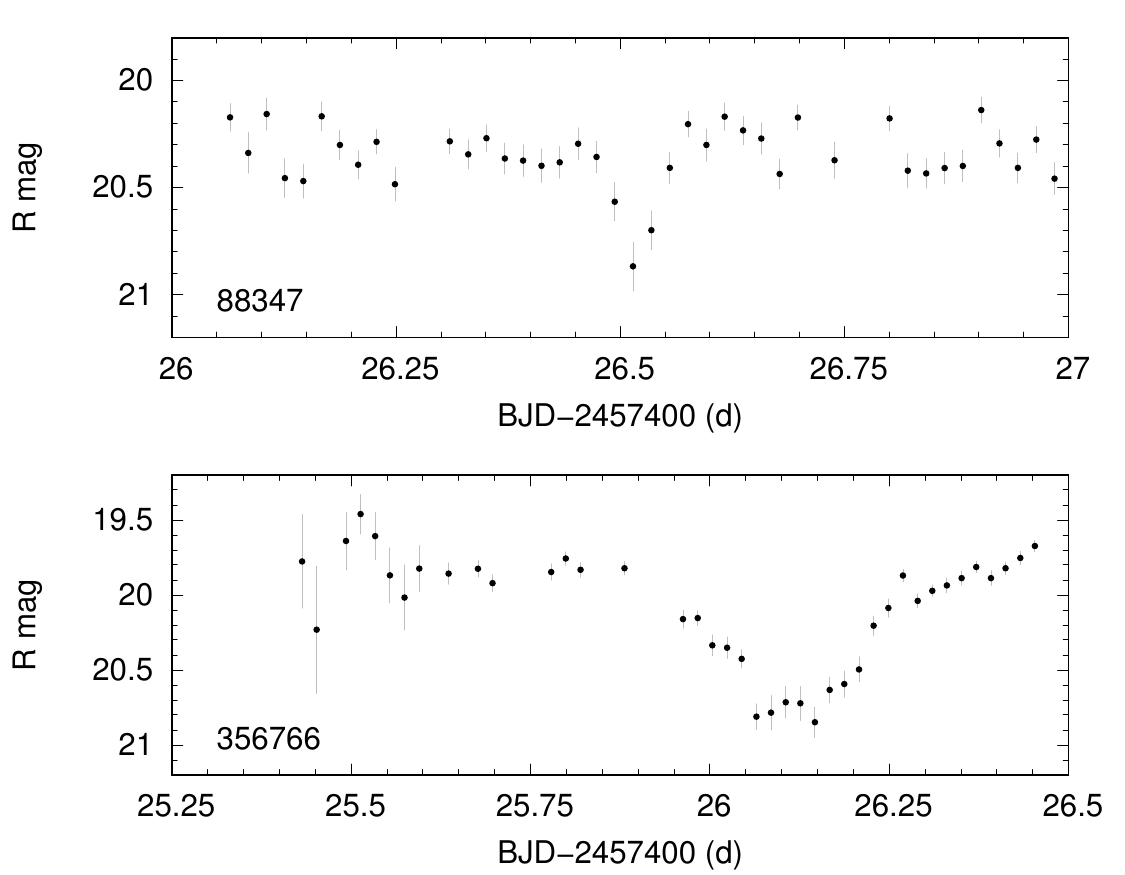}
\caption{Asteroids showing eclipse-like light variations.}
\label{fig:eclipses}
\end{figure}

Many light curves suggest that they originate from contact binary bodies. However, in two cases, we identified distinct eclipse-like features that suggests separated bodies forming asteroid--moon pairs. Unfortunately, as shown in Fig.~\ref{fig:eclipses}, we only captured a single event for (88437) 2000~HL$_{31}$ and (356766) 2004~AF$_{2}$ both, therefore we cannot estimate the orbital periods. The eclipses lasted for $2.3\pm0.2$\,h and $7\pm1$\,h, respectively. The short eclipse of (88437) 2000~HL$_{31}$, compared to the length of the light curve, suggests a wide pair that may be asynchronous, e.g., the main body has a faster rotation period than the orbital period of the moon, but our light curve show no conclusive variations outside the eclipse events \citep[see also][]{scheirich-pravec}. The case of (356766) 2004~AF$_{2}$ is less certain. The first few points of the light curve are affected by one of the diffraction spikes of Uranus, but indicate an extended flat part. The depth of the event (1.0 mag), however, suggests an extensive eclipsing event between similar-sized bodies, a scenario that would favor a close binary. In both cases, follow-up of the systems is desirable.

Some light curves show apparent amplitude changes but in almost all cases these can be traced back either to being very close to the edge of the superstamp or to contamination by the extended halo of Uranus. Such examples are (34402) 2000 RW84 and (83109) that crossed the halo, and 2001 QR$_{240}$ that crossed the saturation column of Uranus, resulting is a distorted first cycles and rejected points in the light curve. However, one asteroid, (94314) 2000 AO$_{165}$ shows slightly changing cycles throughout the observations, including one in the middle with significantly higher amplitude than the rest of the cycles, without approaching the image of Uranus. We added two horizontal lines according the heights of the first minimum and maximum to guide the eye to the bottom panel of Fig.~\ref{fig:threepeaks}. Amplitude changes are observed in the light variations of non-principal axis rotators, and thus the light curve suggests that (94314) 2000~AO$_{165}$ might be a tumbling asteroid \citep[see][]{pravec2005}.

\begin{figure}
%\figurenum{5}
\epsscale{1.18}
\plotone{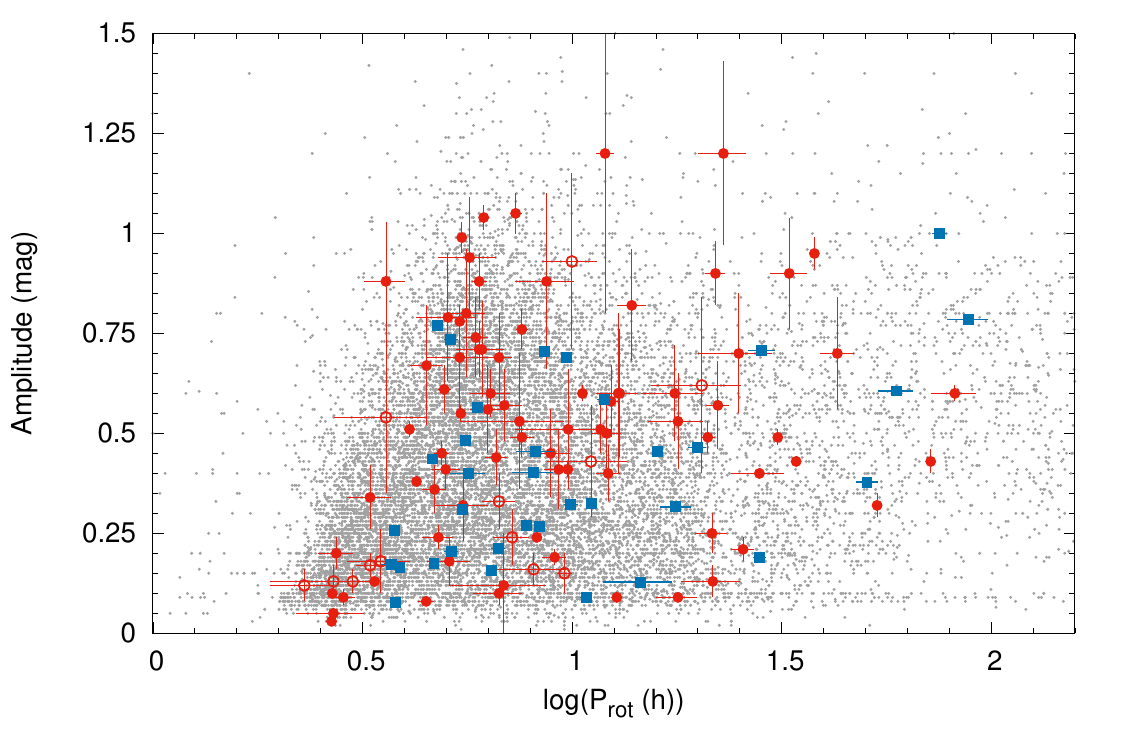}
\caption{Amplitudes versus rotation rates for the MBAs observed by \textit{Kepler}. Blue squares: detections from C0 and C3 \citep{szabo-mba} Red symbols are the objects from this study: empty circles show the uncertain detections. Lower limits are not shown. Small gray points show the distribution of all asteroids with known rotation rates and amplitudes. (Amplitude uncertainties were not listed by \citet{szabo-mba}.) }
\label{fig:rotamp}
\end{figure}

\begin{figure}
%\figurenum{5}
\epsscale{1.18}
\plotone{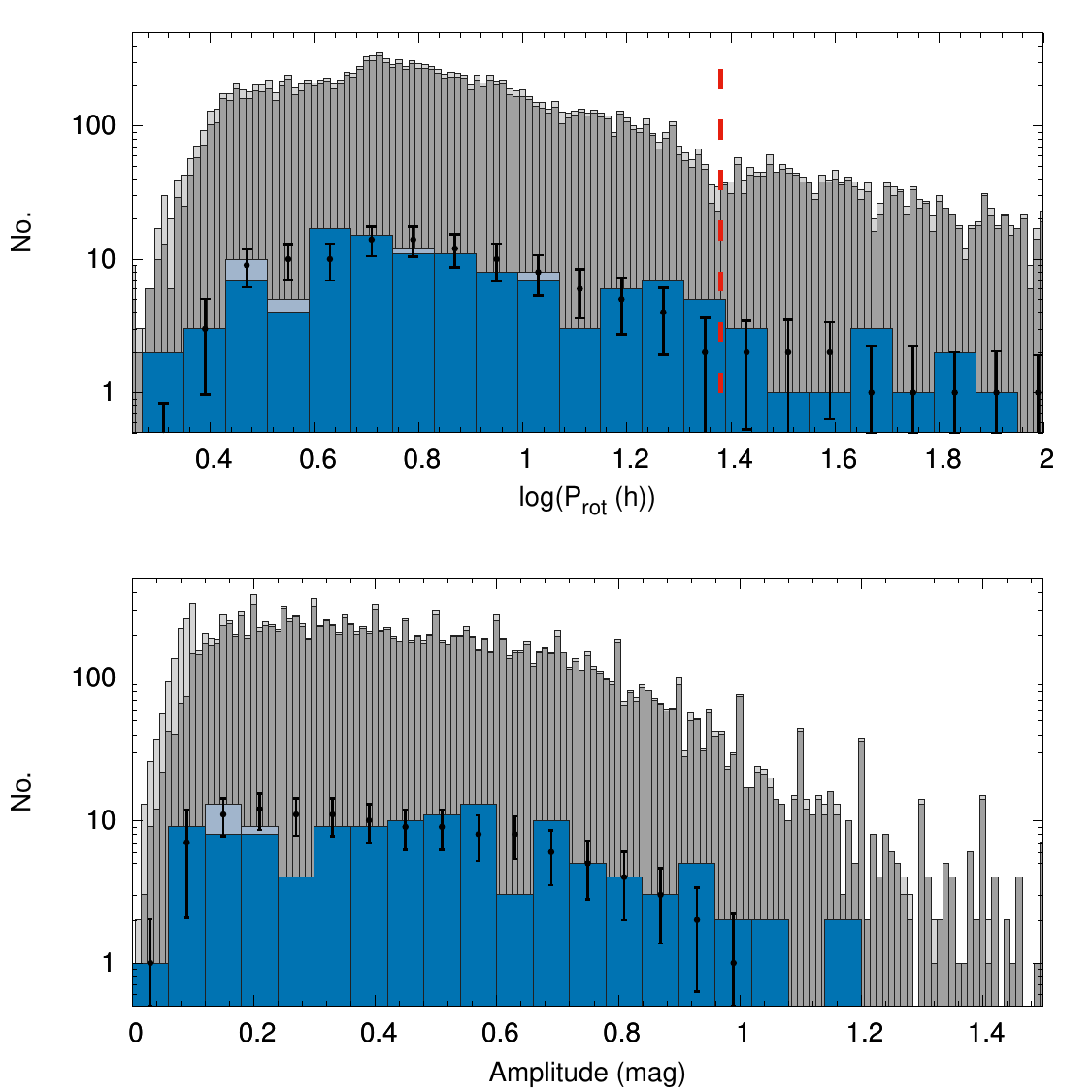}
\caption{Histogram of all observed rotational periods (top) and amplitudes (bottom) from the LCDB (light grey: quality flag U = 1, dark grey: U $> 1$) and the combined K2 sample (blue). Values for asteroids with two possible periods are in light blue. The dashed line marks the 24 hour period. Black points with error bars are simulated results with the same sample size, but based on the LCDB.}
\label{fighist}
\end{figure}

\subsection{Period and amplitude distributions}
We combined our periods and amplitudes with the values obtained by \citet{szabo-mba} and compared the rotation statistics with the sample of all asteroids in the Asteroid Light Curve Database \citep[LCDB,][]{warner-ald} in Figs.~\ref{fig:rotamp} and \ref{fighist}. In Fig.~\ref{fig:rotamp} we compare the K2 detections to the sample of all asteroids with light curves in the LCDB. Data points for 37 asteroids from the K2 fields C0 and C3 (the M35 and the Neptune field, \citealt{szabo-mba}) are plotted in blue, while the new determinations are in red. The overall distribution of K2 asteroids is consistent with the currently known sample, except that there are no very slow rotators in the K2 sample above 100-h period, which can be simply explained by the relative short observation window for K2 asteroids. We separated the LCDB values according to the quality flag U in Fig.~\ref{fighist}: dark grey columns show where periods and amplitudes are either accurate (U $> 2$) or can be considered as good estimates (U $> 1$). We used these values for the statistics below. We also plotted the U = 1 (e.g., uncertain) values in Fig.~\ref{fighist}: only a minority of the sample falls into this category, and they most frequent at the shortest periods and the lowest amplitudes.

In Fig.~\ref{fighist} we plot the period and amplitude distributions of the combined K2 observations separately. Since our sample size is much smaller than the LCDB, instead of a direct comparison, we generated a large number of random test samples from the LCDB, and computed the averages and standard deviations for each bin. These are marked with black symbols and error bars in Fig.~\ref{fighist}.

The LCDB shows a clear deficit of asteroids around the 24-h value (log $P$ = 1.38), as marked in the top panel of Fig.~\ref{fighist}. The K2 sample, in contrast contains no deficit at this value, but rather some excess even in consecutive bins, although the latter is not significant. Nevertheless, our results indicate that the deficit in the literature data is not intrinsic, instead it can be attributed to the inherent difficulties of observing variations on time scales of 12 and 24\,h from the ground. 

Our prior K2 results, based on a smaller sample, hinted that the median rotation rate of the K2 sample could be longer than it is assumed from the (mostly) ground-based observations listed in the LCDB \citep{szabo-mba}. A few years ago the ground-based TALCS survey found an excess of slow rotators after carefully de-biasing their sample \citep{TALCS}. The addition of the Uranus sample tripled the K2 sample size, but the median rotation rate shows only a slight decrease, from 9.83\,h (C0, M35 field) to 9.45\,h (all). This value is still significantly longer than the 7.00\,h median rate of the LCDB: random draws from the database with the K2 sample size yielded a scatter of only $\pm 0.55$ d between the median values. So a 2.45 h difference between the two results cannot be attributed to the low number of objects.

It is of course possible that we missed some fast rotators that have very low amplitudes. Given the long integration time of \textit{Kepler}, very fast rotation can be smeared out and appear as flat light curves. But even at a rotation period of 2.0\,h, observed amplitudes are only decreased by 37\% \citep{szabo-mba}. Comparison with the LCDB actually suggests that the K2 sample does not miss the lowest-amplitude targets, but deficits can be observed in Figs.~\ref{fighist} and \ref{fig:rotamp} at two other areas. Artificially inserting more short-period rotators into the sample to remove these deficits from the histograms can lower the median rotation rate, but it still remains above 8.5 h, indicating that the detection is significant.

\subsection{New detections}

Contrary to our earlier expectations \citep{szabo-mba-initial} we identified a dozen of previously uncatalogized (which can also mean poorly observed, single-opposition) MBAs in the Uranus superstamp with K2. However, according to the new definition of the Minor Planet Center \citep{MPEC-U20} there is a chance that these 21--22~mag detections will not qualify as new discoveries. 

We also searched the superstamp for unknown moons of Uranus. We summed the differential images in 5, 9 and 15 bins to boost the signal of faint, slow-moving objects and to smear out the MBAs. We did not identify any new moons, but discovered one new TNO, a cold Classical Kuiper Belt Object, (506121) 2016~BP$_{81}$ \citep{mpc}. This 22.5~mag distant object was independently detected by the Outer Solar System Origins Survey as well \citep{Bannister-2016}. The detailed light curve analysis of (506121) 2016~BP$_{81}$ will be presented in another paper describing the new K2 observations of TNOs.

\section{Future studies}
\label{sec:out}
Our results show that continuous, space-based photometry can be immensely helpful to gather light curves of Solar System targets, and may uncover biases inherent to ground-based observations. Soon, the TESS mission will launch and provide us, for the first time, continuous, large-scale coverage of the sky, with high photometric accuracy, by collecting full-frame images at every 30 min, for 27 days per observing sector \citep{Ricker2014}. Unfortunately, the mission will have a faint limit of about 16--17 mag, therefore only the bright end of the asteroid population will be covered, but for considerably longer time spans than in our sample. Moreover, the mission will not only deliver continuous light curves, but it will observe opposition surges for the majority of its targets.   

As for the K2 mission, the Uranus field is not the last continuous superstamp observed, but it probably has been the best-suited one for tracking MBAs. A large, continuous area was targeted during Campaign 9 to look for microlensing signals \citep{Zhu2017}, but the immense source density of the Galactic bulge renders it useless for our aims. Other superstamps typically include globular and open clusters, but those are either rather small or crowded, or both. Nevertheless, further, smaller studies can be, in principle, done for the vicinity of M67 (a well-known open cluster), M8 (the Lagoon nebula) and IC 1613 (a nearby dwarf galaxy). Meanwhile, the mission is busy gathering more targeted observations of TNOs, centaurs, comets, Trojans, and even some MBAs, further expanding the unique capabilities of \textit{Kepler} within the field of planetary sciences.

\acknowledgments
The research leading to these results have been supported by the Hungarian Academy of Sciences, by the Hungarian National Research, Development and Innovation Office (NKFIH) grants K-115709, K-119517, PD-116175, K-125015, and GINOP-2.3.2-15-2016-00003, the European Union's Horizon 2020 Research and Innovation Programme, under Grant Agreement No. 687378 and the Lend\"ulet LP2012-31 and LP2014-17 grants of the Hungarian Academy of Sciences. L.~M. was supported by the J\'anos Bolyai Research Scholarship of the Hungarian Academy of Sciences. Funding for the \textit{Kepler} and K2 missions are provided by the NASA Science Mission Directorate. The data presented in this paper were obtained from the Mikulski Archive for Space Telescopes (MAST). STScI is operated by the Association of Universities for Research in Astronomy, Inc., under NASA contract NAS5-26555. Support for MAST for non-HST data is provided by the NASA Office of Space Science via grant NNX09AF08G and by other grants and contracts. The authors thank the hospitality the Veszpr\'em Regional Centre of the Hungarian Academy of Sciences (MTA VEAB), and the University of Pannonia, Veszpr\'em, where parts of this project were carried out.


%% for the paper.  Each keyword is check against the master list during
%% copy editing.  Individual instruments can be provided in parentheses,
%% after the keyword, but they are not verified.

\vspace{5mm}
\facilities{Kepler/K2 \citep{howell-k2}}

\software{FITSH \citep{fitsh}, EPHEMD, Period04 \citep{period04,period04-source}, NumPy \citep{numpy}, R, gnuplot}

\bibliography{uranus_asteroids_R2}

\appendix
\section{Light curves}
\label{sec:app}
\begin{figure*}[hb!]
\epsscale{1.15}
\plotone{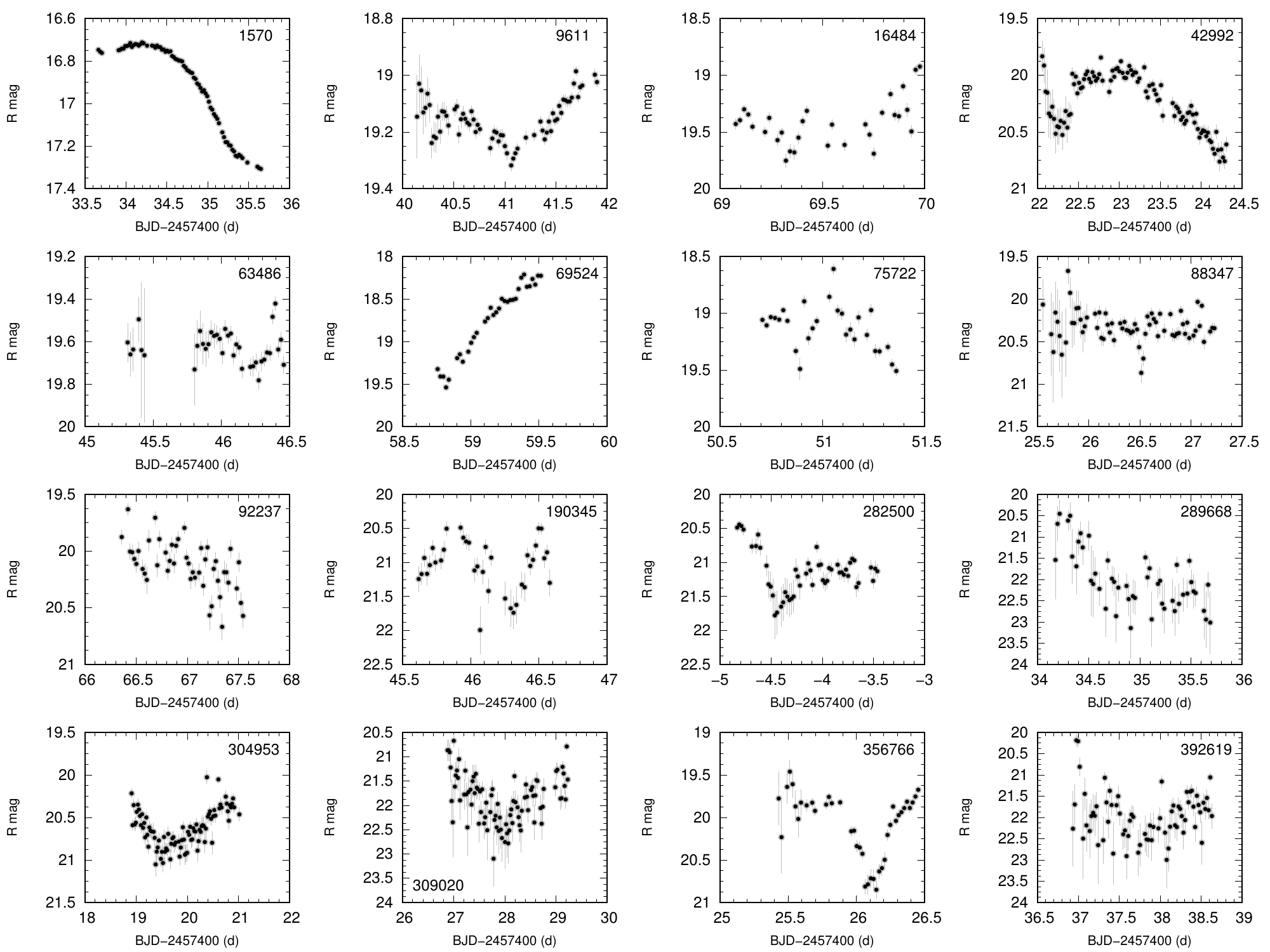}
\caption{Asteroid light curves showing slow variations, from Campaign 8 of the K2 mission. }
\label{fig:longper}
\end{figure*}

\begin{figure*}[hb!]
\epsscale{1.06}
\plotone{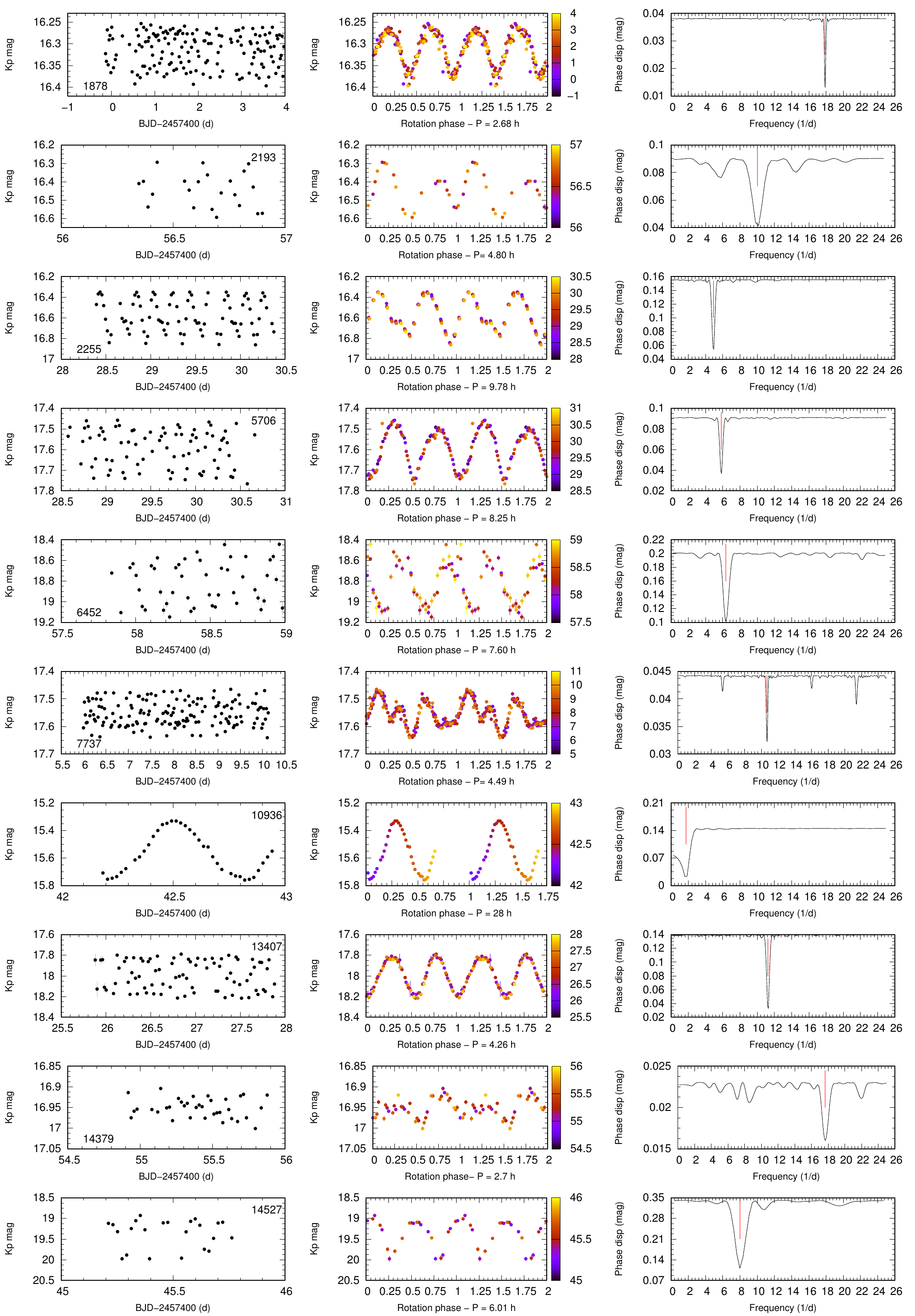}
\caption{Asteroid light curves from Campaign 8 of the K2 mission.  }
\label{fig:ast01}
\end{figure*}

\begin{figure*}[hb!]
\epsscale{1.06}
\plotone{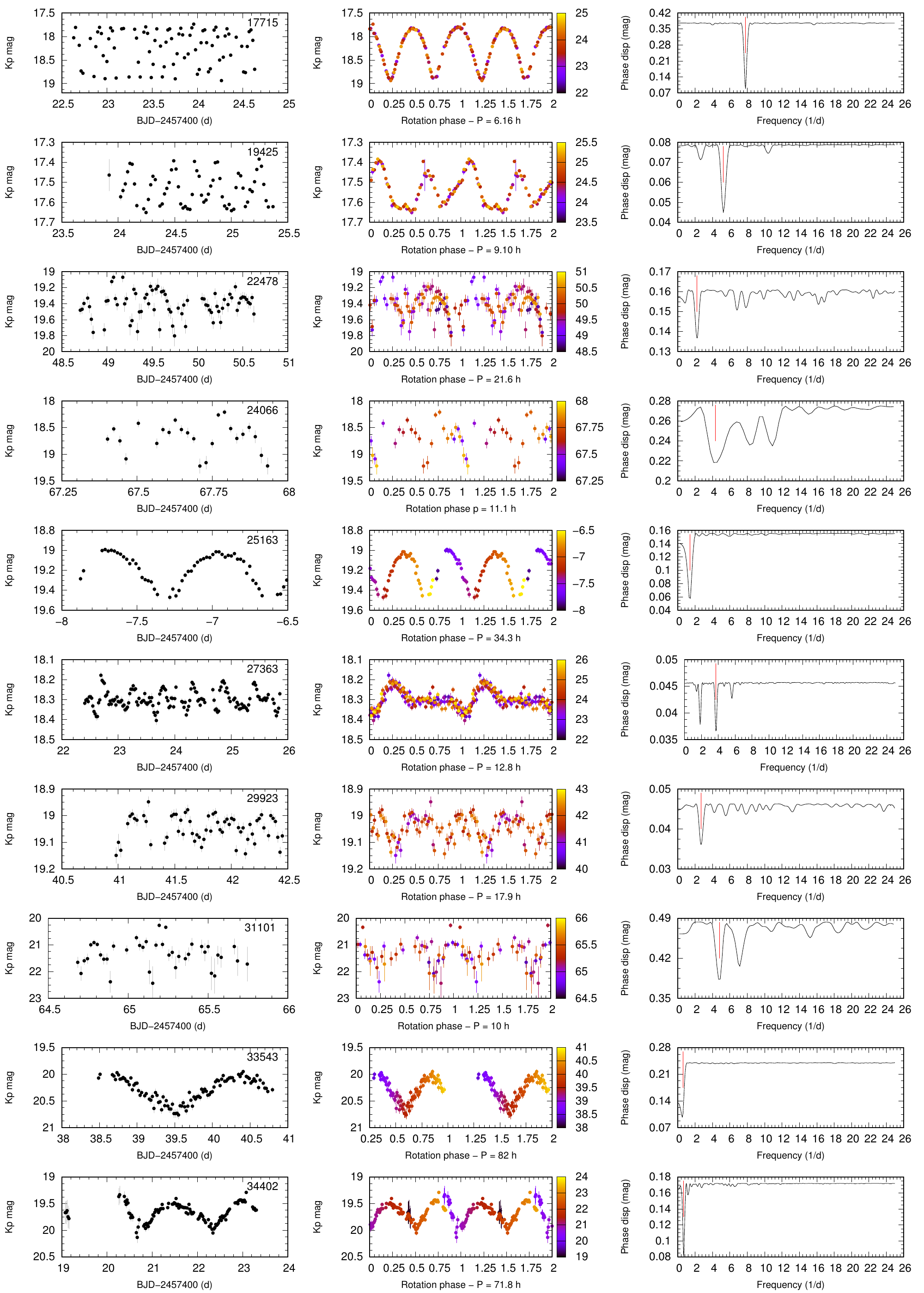}
\caption{Asteroid light curves from Campaign 8 of the K2 mission (continued).  }
\label{fig:ast02}
\end{figure*}

\begin{figure*}[hb!]
\epsscale{1.06}
\plotone{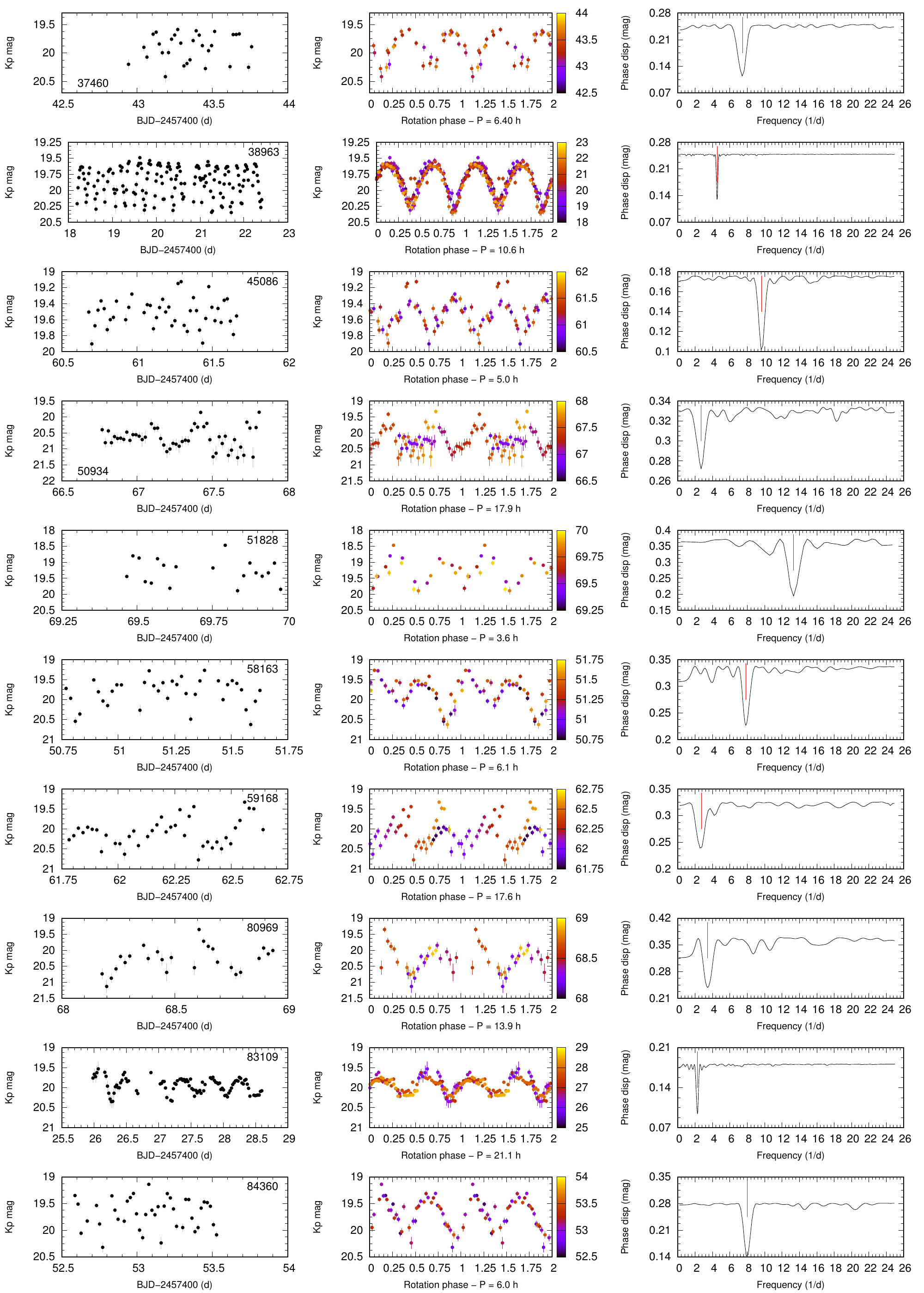}
\caption{Asteroid light curves from Campaign 8 of the K2 mission (continued).  }
\label{fig:ast03}
\end{figure*}

\begin{figure*}[hb!]
\epsscale{1.06}
\plotone{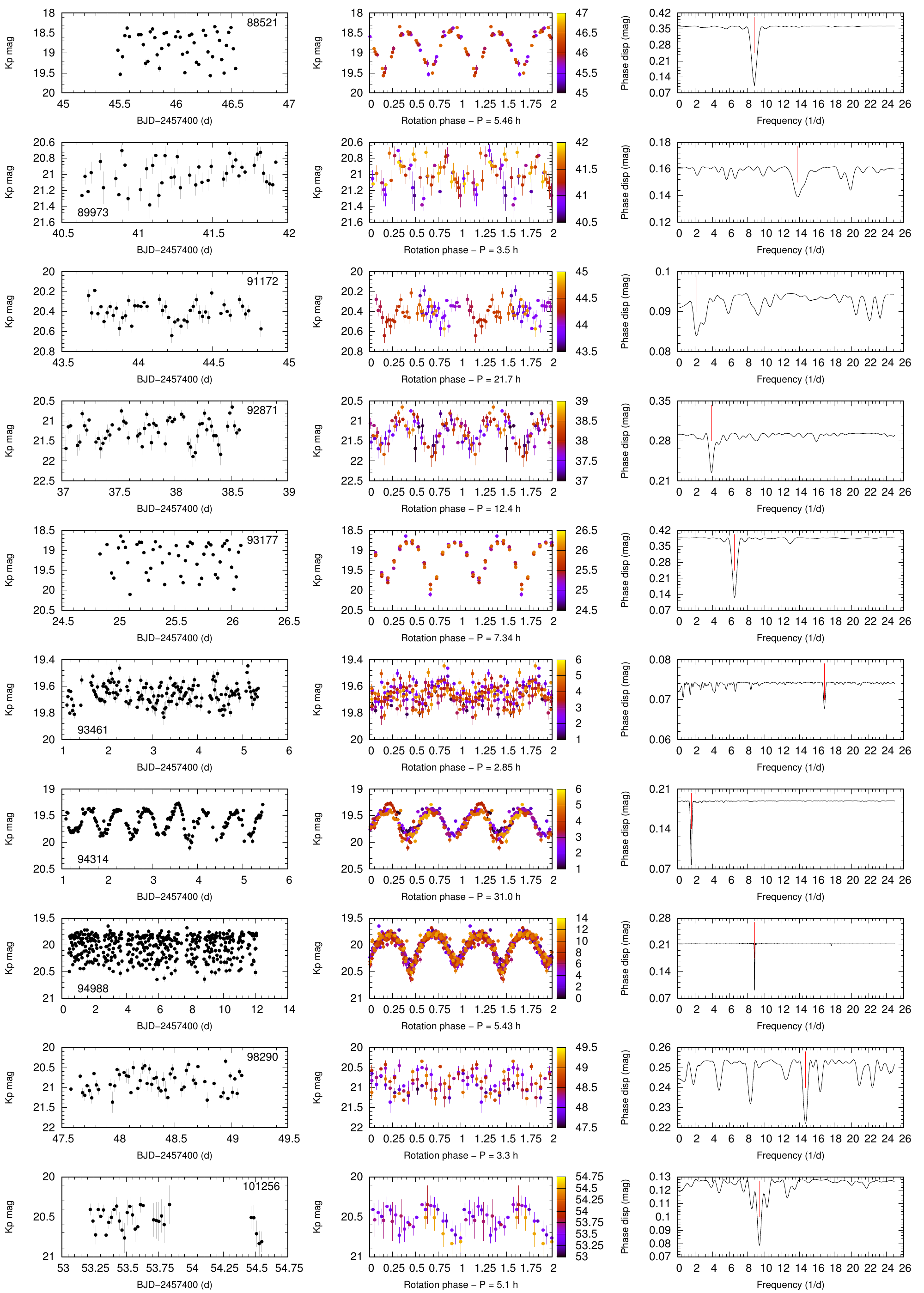}
\caption{Asteroid light curves from Campaign 8 of the K2 mission (continued).  }
\label{fig:ast04}
\end{figure*}

\begin{figure*}[hb!]
\epsscale{1.06}
\plotone{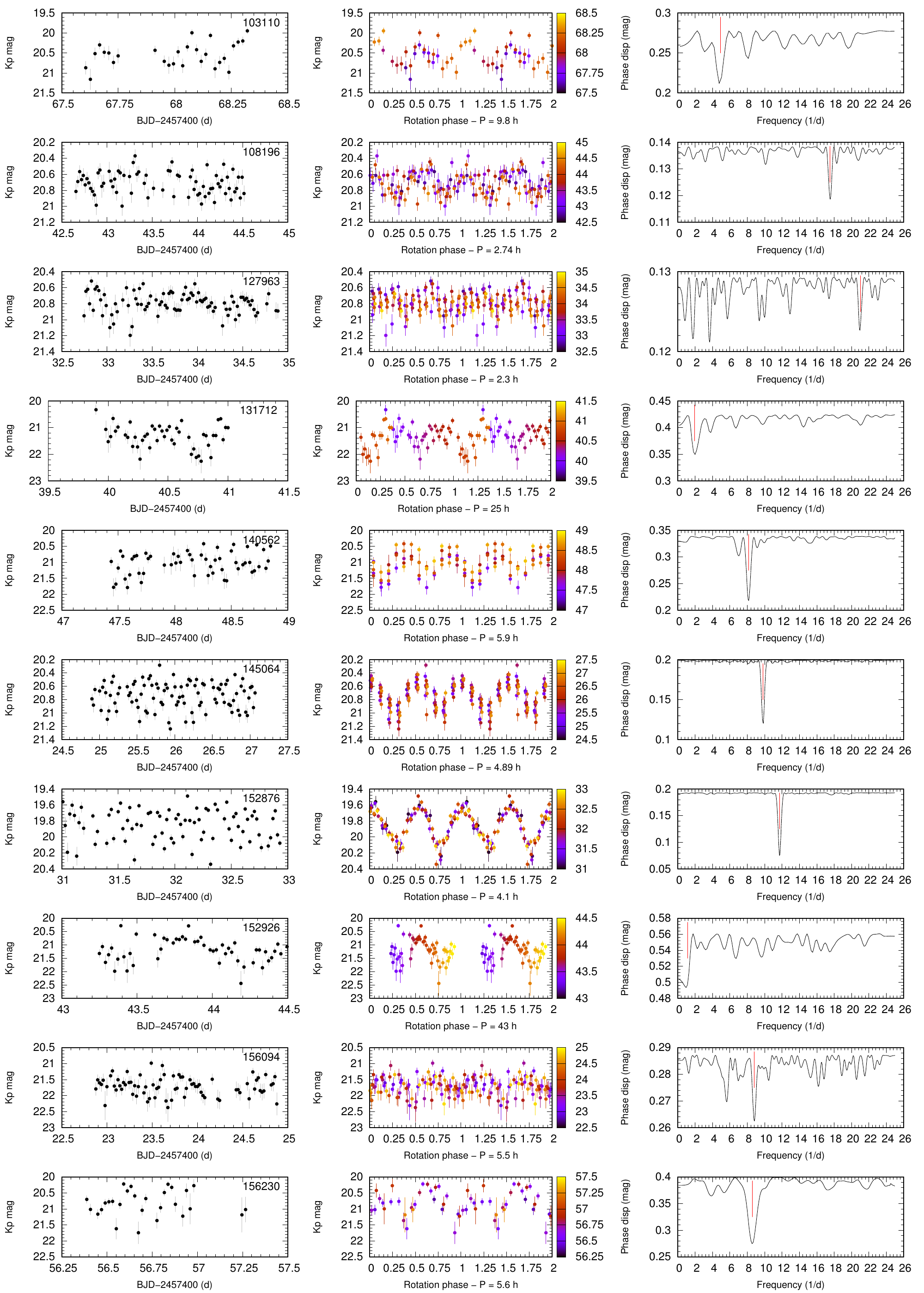}
\caption{Asteroid light curves from Campaign 8 of the K2 mission (continued).  }
\label{fig:ast05}
\end{figure*}

\begin{figure*}[hb!]
\epsscale{1.06}
\plotone{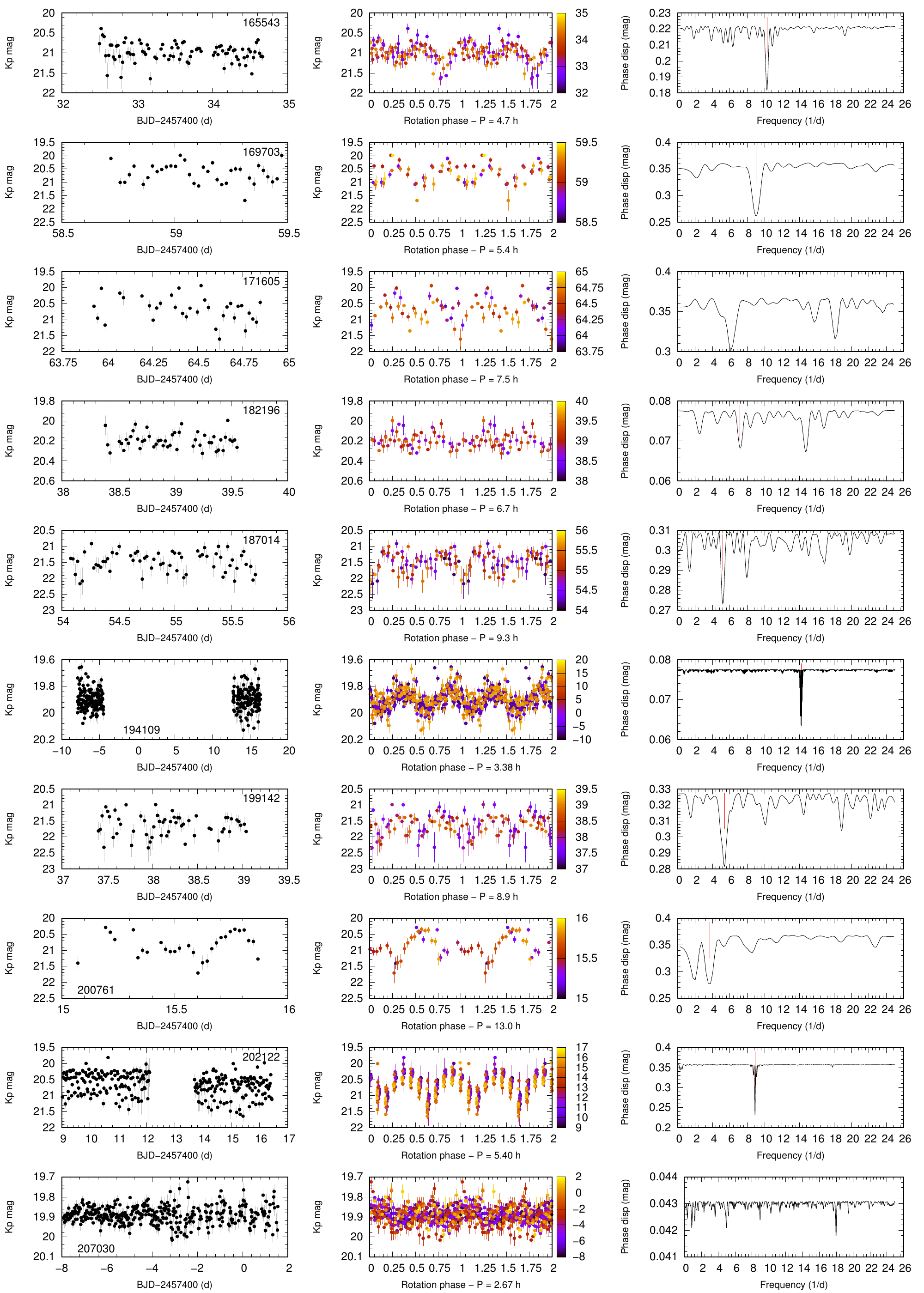}
\caption{Asteroid light curves from Campaign 8 of the K2 mission (continued).  }
\label{fig:ast06}
\end{figure*}

\begin{figure*}[hb!]
\epsscale{1.06}
\plotone{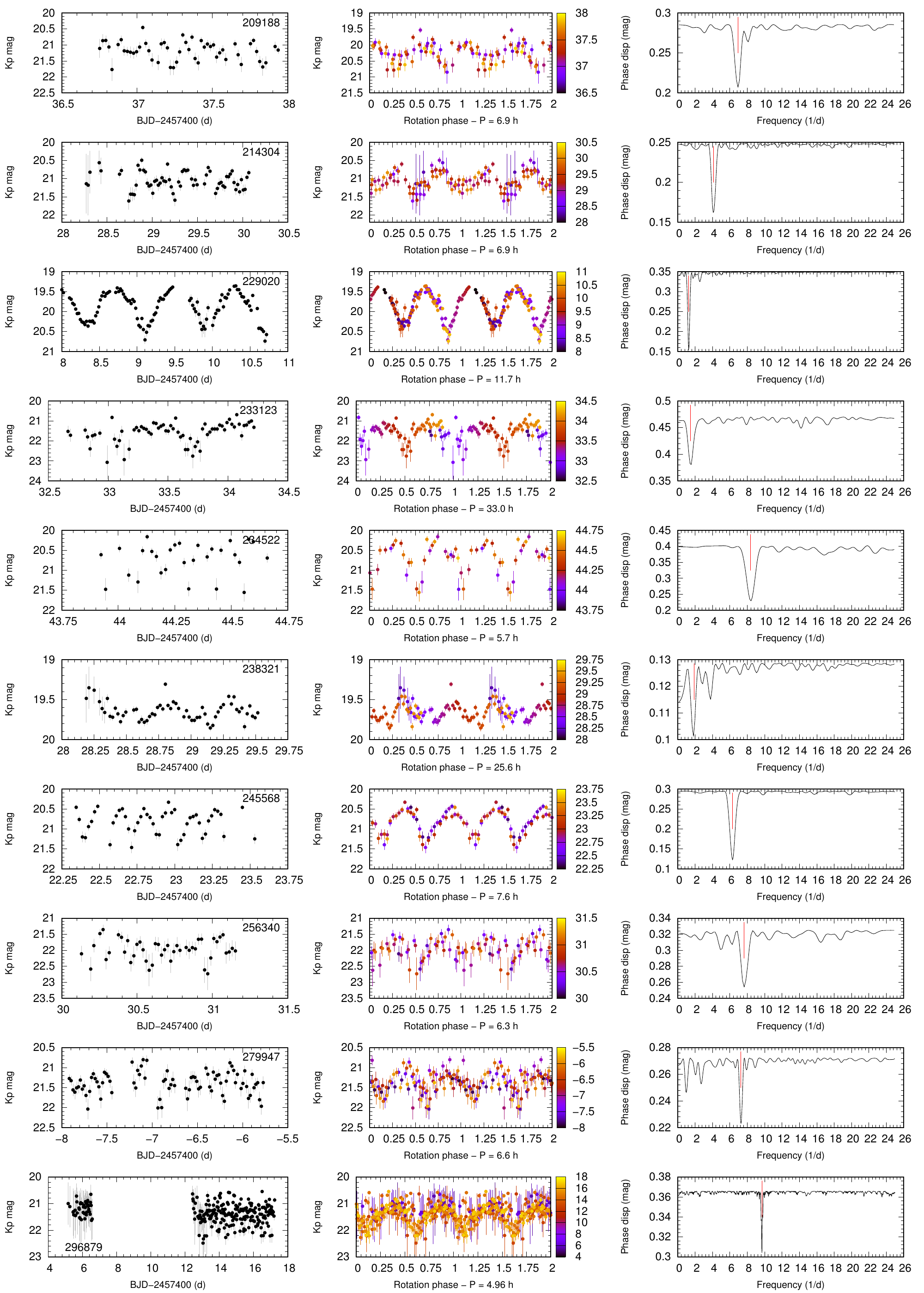}
\caption{Asteroid light curves from Campaign 8 of the K2 mission (continued).  }
\label{fig:ast07}
\end{figure*}

\begin{figure*}[hb!]
\epsscale{1.06}
\plotone{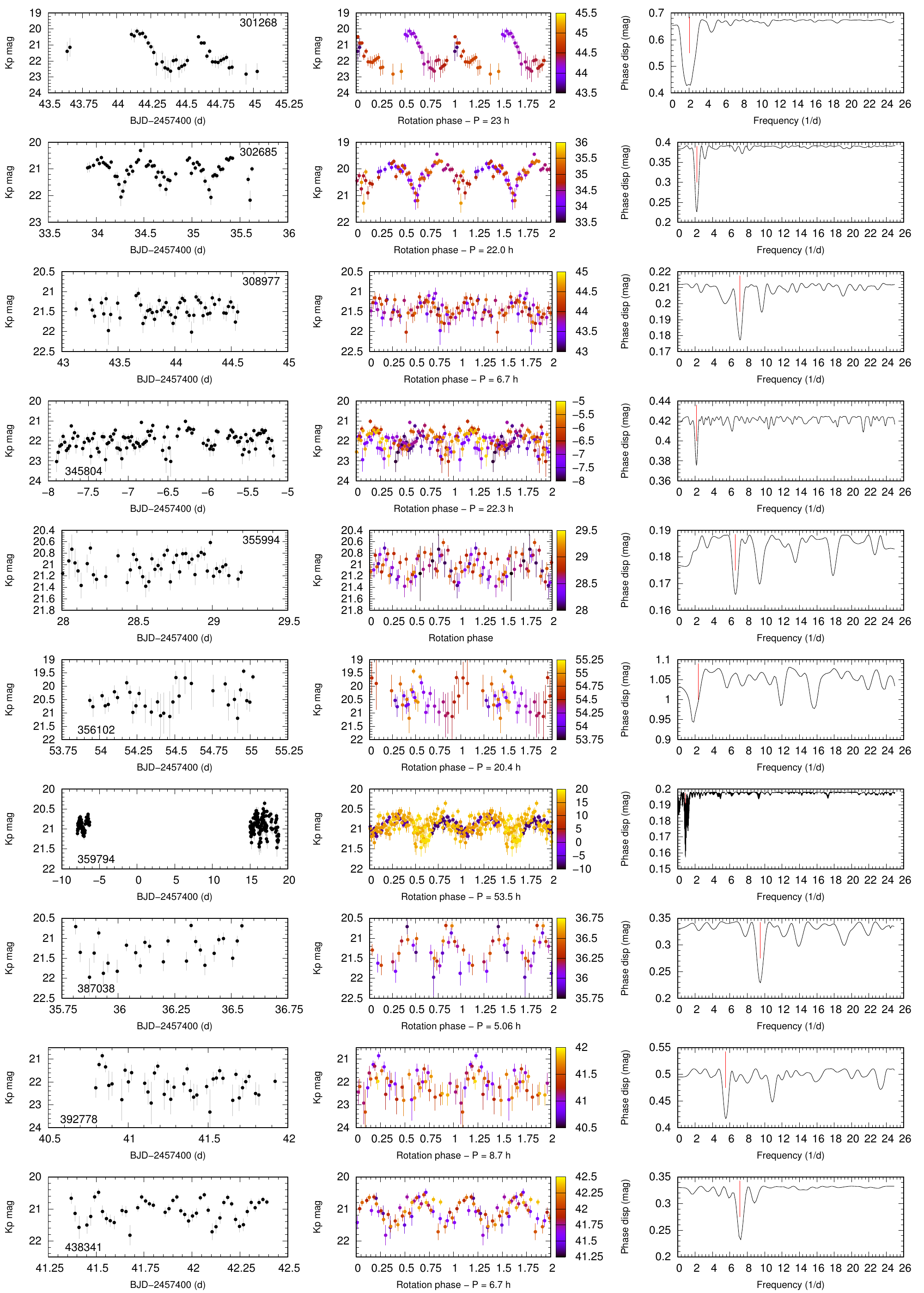}
\caption{Asteroid light curves from Campaign 8 of the K2 mission (continued).  }
\label{fig:ast08}
\end{figure*}

\begin{figure*}[hb!]
\epsscale{1.06}
\plotone{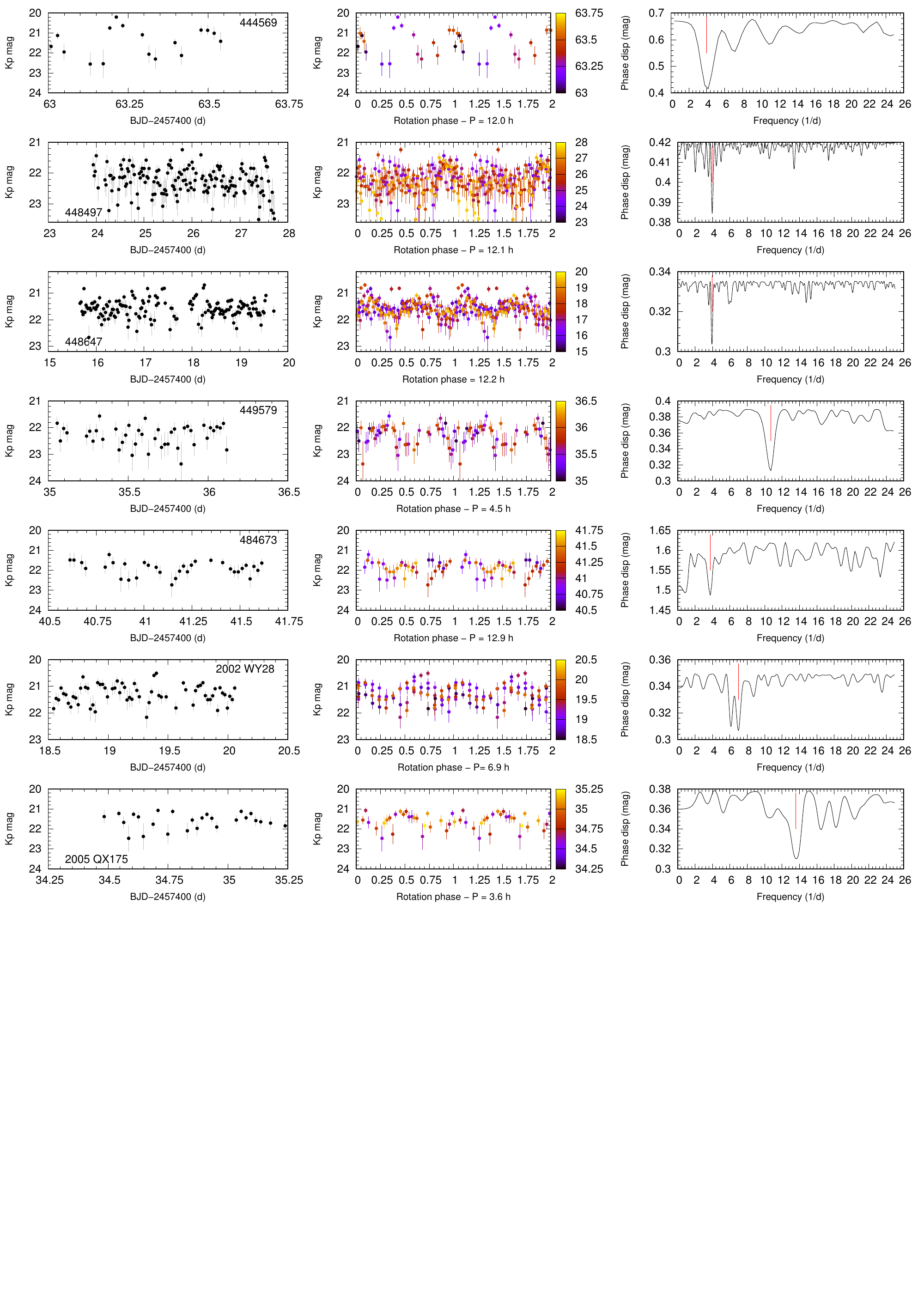}
\caption{Asteroid light curves from Campaign 8 of the K2 mission (continued).  }
\label{fig:ast09}
\end{figure*}

\begin{figure*}[hb!]
\epsscale{1.06}
\plotone{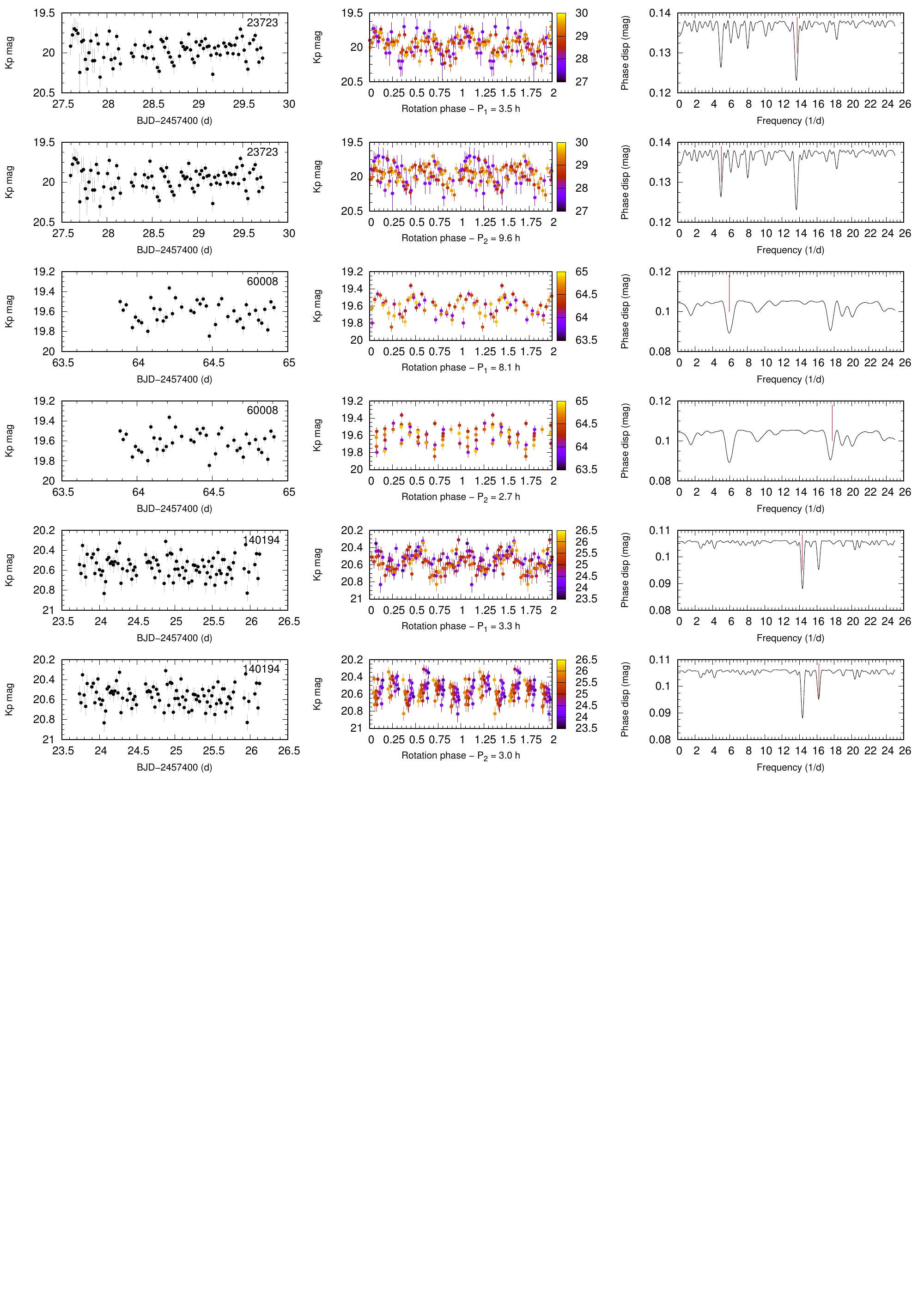}
\caption{Asteroid light curves from Campaign 8 of the K2 mission with two possible periods.  }
\label{fig:ast10}
\end{figure*}

Below we display the light curves of all asteroids where we detected rotational variations. Figure~\ref{fig:longper} shows the light curves with incomplete cycles. Figures \ref{fig:ast01}--\ref{fig:ast10} present the asteroid detections where we were able to determine the periods. The left column shows the light curves themselves; the middle column are the phase curves, folded with the rotation period while the color scale represents time (BJD-2457400). The right column displays the residual dispersion plots, where the red line marks the single-peak frequency ($2/P_{\rm rot}$). In the last page we show the solutions for the three asteroids where two possible periods have been identified. Note that the dispersion of the residual drops at the characteristic frequencies, therefore the minima are sought, contrary to Fourier transforms or periodograms. 

\listofchanges

\end{document}